# Materials Informatics: Emergence To Autonomous Discovery In The Age Of AI

Turab Lookman[1,2,*], YuJie Liu[1], Zhibin Gao[1,†]

[1]State Key Laboratory for Mechanical Behavior of Materials, State Key Laboratory of Porous Metal Materials, School of Materials Science and Engineering, Xi'an Jiaotong University, Xi'an 710049, China
[2]AiMaterial Research, Santa Fe, New Mexico, USA 87501
[*]E-mail: turablookman@gmail.com
[†]E-mail: zhibin.gao@xjtu.edu.cn

We provide a perspective on the evolution of materials informatics, tracing its conceptual roots to foundational ideas in physics and information theory and its maturation through the integration of machine learning and artificial intelligence (AI). Early contributions from Chelikowsky, Phillips, and Bhadeshia laid the groundwork for what has become a transformative approach to materials discovery. The U.S. Materials Genome Initiative catalyzed a surge in activity, and the period from 2014–2016 marked the first impactful applications of machine learning to materials problems.

Since then, the field has seen rapid advances, particularly with the advent of deep learning and transformer-based large language models (LLMs), which now underpin tools for property prediction, synthesis planning, and inverse design. We present the subject not as a collection of tools, but as an evolving research ecosystem – tracing the historical context, current situation and future development direction of the field. We review key methodologies—including approaches for sequential design, such as Bayesian Optimization and Reinforcement Learning, and transformers —highlighting their role in accelerating discovery and automating experimentation with growing efforts in building autonomous self-driving laboratories. We discuss common pitfalls, issues and solutions associated with LLMs and Bayesian Optimization as applied to the study of specific materials systems. Given the costs of pre-trained AI models for specific materials, we consider the merits of specialist LLMs versus today's state-of-the-art generalists.

Finally, we assess emerging challenges and the potential for AI to evolve from a predictive tool into a collaborative partner in research. We argue that with advances in active learning, uncertainty quantification, and retrieval-augmented generation, a new era of autonomous materials science is within reach—one in which the human is increasingly taken out of the loop.

Bio：
Turab Lookman obtained his Ph.D. from Kings College, University of London, and held appointments at Western University and the University of Toronto in Canada until 1999, and at Los Alamos National Laboratory till 2019. He was elected Fellow of the American Physical Society (APS) in 2012 and a Laboratory Fellow at Los Alamos National Laboratory in 2018. His interests and expertise lie in hard and soft materials science and condensed matter physics, applied mathematics, and computational methods. His work on information directed approaches to materials discovery started in 2012 when he proposed the use of an active learning approach based on Bayesian Global Optimization. The concept was funded by LANL/DOE in 2013 to investigate how ML tools may be applied to accelerate materials discovery. The work led to the synthesis of materials with targeted response. He has published over 450 papers, has ~ 20K citations with an h index of 66 (Google Scholar).

Yujie Liu is a Ph.D. in the School of Materials Science and Engineering at Xi'an Jiaotong University. They obtained their bachelor's degree from the School of Materials Science and Engineering at Wuhan University of Technology. His research background is in the application of machine learning combined with domain knowledge for the prediction of lattice thermal conductivity. He has published papers in Acta Physica Sinica and the Journal of Materials Informatics. Currently, their research focuses on the application of machine learning and large language models for the prediction and design of properties in metallic materials.

Zhibin Gao is an associate professor in the School of Materials Science and Engineering at Xi'an Jiaotong University and a recipient of the prestigious Shaanxi Province Young Talent award. He earned his Ph.D. from Tongji University and completed his postdoctoral training at the National University of Singapore. His research is highly interdisciplinary, focusing on condensed matter physics, computational materials science, machine learning, and the application of large language models. To date, he has authored 84 SCI-indexed publications in leading journals, including *Energy & Environmental Science*, *Advanced Materials*, *Advanced Science, Advanced Functional Materials*, *ACS Nano*, *Nano Letters*, and the *Physical Review* series. His work has garnered over 2,500 citations, resulting in an h-index of 25. In addition to his publications, he holds two software copyrights.

# 1 Introduction: Origins and Paradigm Shifts

Materials informatics is an emerging area fusing aspects of computer science and machine learning with statistical inference and materials science. There has been considerable progress in the last few years in the application of tools from machine learning and artificial intelligence (AI) to materials science[1–10]. AI in its strict sense is any application that can perform a task without human intervention. Within AI is Machine Learning and within it is Deep Learning and much of the current euphoria about AI relates to the impact of neural network-based architectures, such as Transformers, on our daily lives. The impact on materials science has led to a paradigm shift in the way materials science is studied, and now many transformer based large language models (LLMs) have been developed for many subfields within materials science devoted to the prediction of materials properties[2,11–13]. In addition, autonomous self-driving laboratories are rapidly expanding everywhere to carry out experiments on the fly controlled by active learning algorithms that can guide discovery[14–17].

It is a rapidly emerging interdisciplinary field, integrating principles from computer science, machine learning, and statistical inference with core materials science[18–32]. Whereas AI broadly encompasses systems performing tasks without human intervention, ML is a key subset, and deep learning (DL), with its neural network architectures, such as Transformers, drives much of the current AI revolution. This has profoundly impacted materials science, ushering in a paradigm shift. Transformer-based large language models (LLMs) are now prevalent across various materials subfields, particularly for predicting material properties. Concurrently, autonomous self-driving laboratories are rapidly expanding globally, leveraging active learning algorithms to guide on-the-fly experimental discovery.

We trace below the historical emergence of materials informatics (Figure 1), from foundational concepts by Schrödinger and Shannon, to seminal contributions by Chelikowsky and Phillips in 1976, and more recent work by Bhadeshia in the 1990s. The U.S. Materials Genome Initiative (MGI), launched in 2011, significantly catalyzed activity, with the first notable applications of machine learning to materials discovery emerging around 2014-2016. Since then, the field has experienced transformative growth, largely driven by deep learning. We will review key data-driven tools employed in MGI, including neural networks, Bayesian optimization, reinforcement learning (RL), and the role of Transformers and LLMs, examining their influence on materials discovery, both through traditional synthesis and increasingly through autonomous approaches. Although there have been a number of reviews on different aspects of AI-powered materials informatics, many of which we cite, our focus is on presenting the subject not merely as a collection of tools, but as an evolving research ecosystem – tracing the historical context, current situation and future development direction of the field. Moreover, we discuss the common pitfalls, issues and solutions associated with tools, such as LLMs and Bayesian Optimization, when applied to the study of specific materials systems. In particular, we examine matters of accuracy and reproducibility. Given the high costs associated with creating pre-trained AI models for specific materials, we assess the merits of specialist LLMs versus today's state-of-the-art generalists, such as DeepSeek and Gemini, by comparing their predictions for alloys. We conclude by addressing the challenges that persist in the journey toward achieving fully human-out-of-the-loop materials discovery[33].

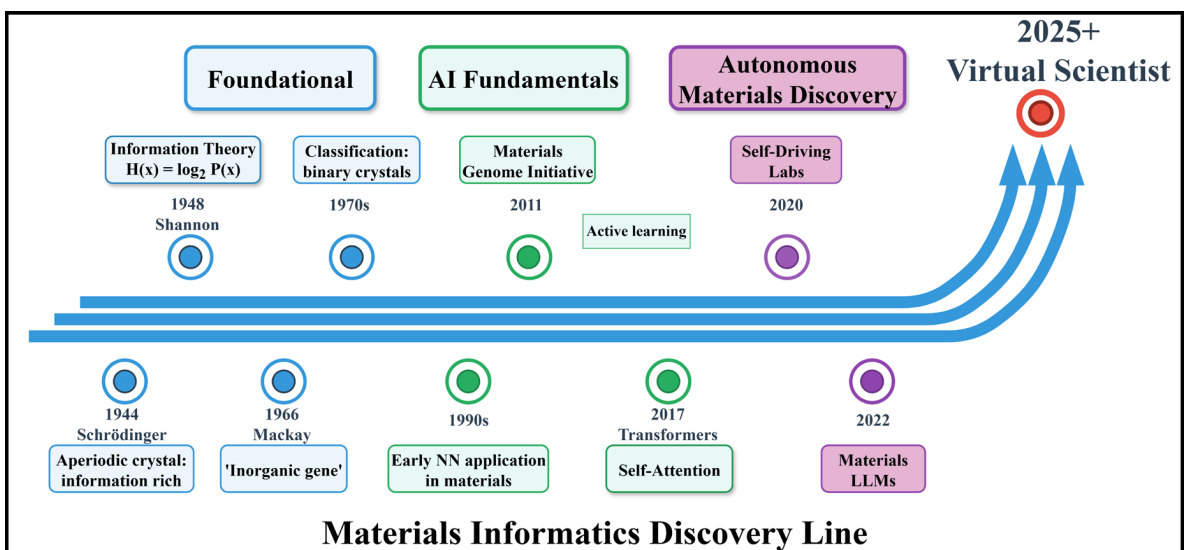

Figure 1. The evolving Materials Informatics research ecosystem.

*1.1 Information, Aperiodic Crystals, and the Birth of Materials Informatics*

In his 1944 book, What is Life?[34], Erwin Schrödinger suggested the concept of an aperiodic crystal as a means of passing hereditary traits of individuals from parent to offspring. His underlying idea, long before the structure of DNA was solved, was that some structure or vehicle had to be present that lacked periodicity so that it was not information poor to conduct the hereditary process. Following Shannon's work on communications[35], we now appreciate that the presence of information constitutes surprise or unpredictability so that if a set of messages, x occur with probability P(x), then if self information, $H(x) = -\log_2 P(x)$, predictable messages with $P(x) = 1$ will yield no information or surprise.

These ideas were incorporated in the work of the crystallographer A. MacKay[36] of Birbeck College who considered crystallography to apply to all structures that showed some form of organization, far beyond periodic structures. He suggested "a crystal is a structure the description of which is much smaller than the structure itself and this view leads to the consideration of structures as carriers of information and on to wider concerns with growth, form, morphogenesis and life itself". He thus had in mind the notion of an inorganic gene or descriptors in the context of materials genome.

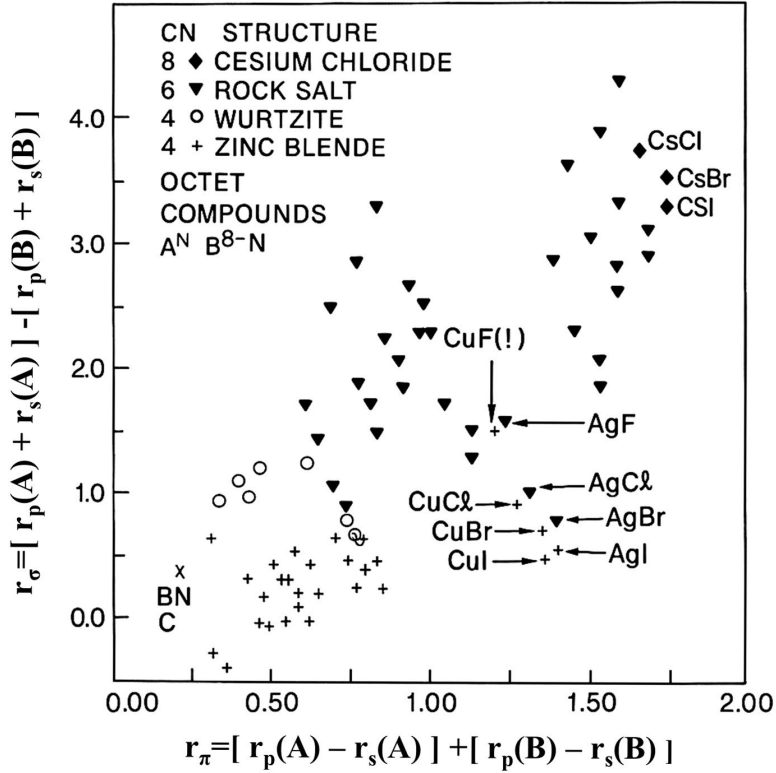

Figure 2 The St. John—Bloch plot for 79 binary octet crystals $A^N B^{8-N}$ by Chelikowsky and Philips[37]. The classification precision achieved manually by these early pioneers by appropriate selection of atomic descriptors was in excess of 85%. Reproduced with permission. [37] Copyright 1978, APS Journals.

Some of the earliest studies on what we identify now as Materials Informatics, date back almost sixty years to the classic problem of separating binary solids into their crystal structures, namely, rocksalt, zincblende, wurtzite, cesium chloride, and diamond. Several authors[28-30] (Figure 2) had recognized the importance of materials features that classify the binary octet solids. St. John and Bloch[39] had proposed as features the x and y co-ordinates of the symmetric combinations. $r_\sigma = |(r_p^A + r_s^A) + (r_p^B + r_s^B)|$ and $r_\pi = |(r_p^A - r_s^A) + (r_p^B - r_s^B)|$ of the s and p orbital dependent radii of the A and B atoms estimated from early pseudo-potential first principles calculations. They applied their scheme to a set of 63 $A^N B^{8-N}$ compounds. Their plot was subsequently generalized by Chelikowsky and Philips[37] by including the noble elements Cu, Ag and Au. Figure 2 shows the St. John and Bloch plot by Chelikowsky and Philips[37] for 79 binary octets. This was subsequently applied to all AB materials with the symmetric combinations of ionic radii but with $r_A$ and $r_B$ computed by a different pseudo-potential[38], and the 574 AB materials then known were classified into 34 crystal structures. Today we recognize this as a classification problem, but at the time physical insight played a key role in identifying Pauling's electronegativity scale, average principal quantum number and orbital radii of the elements as descriptors or features that enabled the structures of octet subset of AB solids to be separated in a 2D structural plot. We are now able to utilize machine learning methods and invoke more refined features, however, what is remarkable is that early studies were able to manually achieve accuracies more than 85% in classifying available data. Even as early as the 1970's, Chelikowsky and Philips[37] using data generated from early first principles calculations to find semiconductors, recognized that the structural energy differences between the solids were often too small to be calculated by methods and computers available at the time, and that information theoretic concepts were necessary to learn from the structural data available then to extract rules for chemical bonding.

*1.2 Early Neural Networks in Materials Research*

The application of neural network (NN) methods to study the mechanical behavior of materials also has a distinguished history that dates back thirty years. Ghaboussi et al.[40] in 1991 recognized that neural networks provided an alternative paradigm to the development of physical models of materials that directly made use of available experimental data. Their capability to self-organize by learning from the data and making predictions provided a means of making progress on complex materials, such as composites, where few physical principles and models were available. Their pioneering application was to use a back propagation neural network with six input and two output nodes with two hidden layers of forty nodes each to study the biaxial behavior of plain concrete. They had good experimental data on stress-controlled tests and since the problem was path dependent, they trained the network to predict strain increments given the current strain, stress and stress increments. Thus, the inputs were two stresses, two strains and two stress increments and the outputs were the two strain increments. Figure 3a shows the neural network used from this pioneering study applied to concrete cyclic loading data. The network was initially trained on the first four cycles and tested on the fifth cycle. Comparison to experiments shows that the model captures the trend. It is extraordinary that the type of problems that are being studied today with neural networks were first investigated three decades ago.

Neural networks in materials science, including phase transformations, multicomponent alloys and steels are closely associated with the work of Prof. H.K.D.H Bhadeshia. He considered nonlinear networks as straight forward generalizations of linear models involving variables (Figure 3b), such as compositions of elements, and unknown parameters fitted to data[41]. His interests were stimulated in 1992 when he did not understand a presentation at a conference and approached D. MacKay at Cambridge, who worked in the information sciences field. This led to a collaboration on the application of neural networks to the impact toughness of C–Mn Steel Arc–Welds[42]. Using a network of fourteen input nodes and one output node, the authors studied the influence of variables such as alloy composition and percentage of primary and secondary microstructure, temperature, yield strength on the toughness. They varied the complexity of the network by varying the number of nodes in the hidden layer from two to seven and found that four nodes gave optimal results based on a Bayesian analysis in terms of the *evidence,* the probability of a model. The model gave very reasonable results and was used to study effects that could not be studied experimentally, such as whether acicular ferrite was a better microstructure than Widmanstätten ferrite. The success of this work led Bhadeshia and Mackay to collaborate on many other problems over a twenty-four-year period.

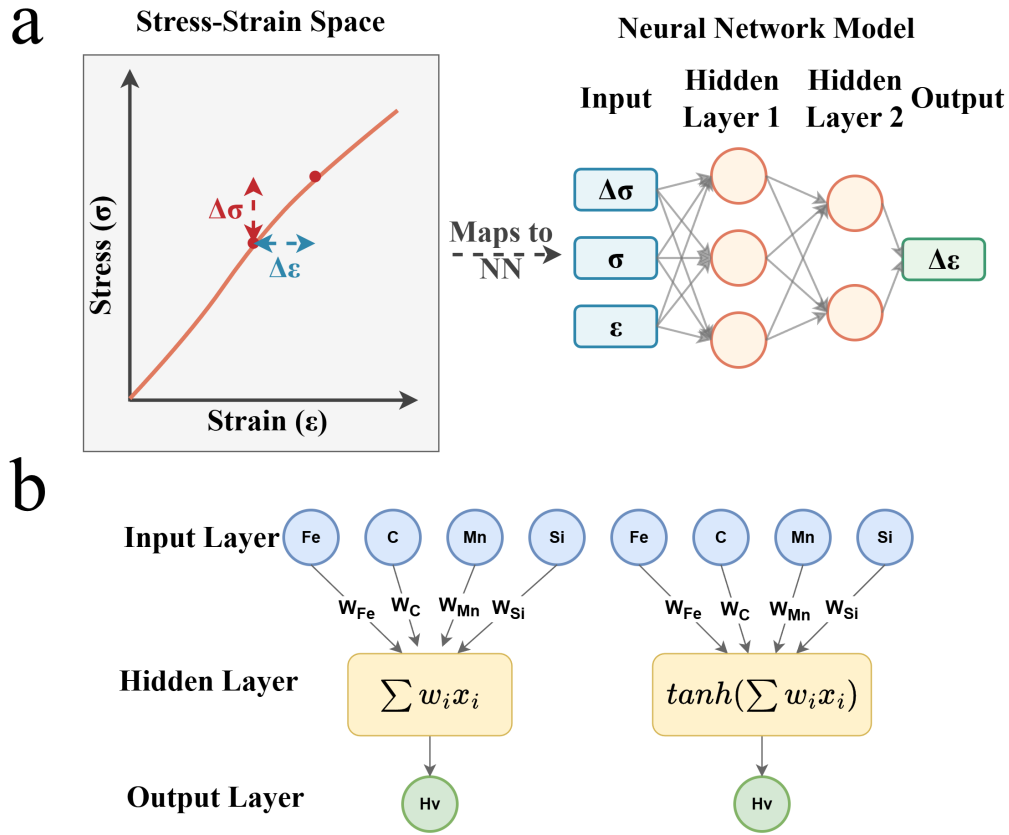

Figure 3 a) A neural network model for Concrete under biaxial stress loading studied in 1991 by Ghaboussi et al[40]. The inputs are the stresses, strains and stress increments and the output are the strain increments. The hidden layers contained 40 nodes each. b) Nonlinear network as generalization of a linear network or model with composition inputs (variables) and parameters used in studies on alloys dating back to the 1990's[41].

*1.3 From Trial-and-Error to early Materials Discovery*

Before the launch of the Materials Genome Initiative (MGI) in 2011, materials discovery and design were predominantly empirical and computationally limited. MGI transformed materials science by integrating high-throughput computations, data-driven methods, and collaborative databases to accelerate material discovery, aiming to reduce the time it took to bring materials to market by half. The initiative spanned the entire process, from materials discovery and property optimization to systems engineering deployment, with different agencies determining the specifics of its implementation. MGI catalyzed a shift that shortened the timeline for discovering new materials from over two decades to under ten years[43].

Historically, materials discovery was often described as "Edisonian," relying heavily on trial-and-error experimentation. Around 90% of industrial R&D in materials, including aerospace alloys and semiconductors, depended on physical experiments and expert intuition. This approach used heuristics to synthesize alloys, ceramics, and polymers through methods like furnaces and sputtering, with key experimental tools such as X-ray diffraction (XRD), electron microscopy (SEM/TEM), and mechanical testing. While this approach led to significant developments like steels based on empirical phase diagrams and superalloys optimized via heat treatments, the process was slow, taking 10 to 30 years from lab to market, and hindered by computational limitations.

In the late 1990s and early 2000s, combinatorial materials science emerged, where automated systems created gradient composition libraries for thin films. These were screened optically and mechanically, leading to breakthroughs in catalysis, though this method was limited to thin films. High-throughput approaches, while groundbreaking, were costly (over $1 million per new alloy) and slow, often taking several years per material. Major breakthroughs, such as the discovery of high-entropy alloys in 2004, were often serendipitous[44].

On the computational front, early theoretical models, such as CALPHAD (CALculation of PHAse Diagrams), predicted material behavior based on thermodynamic data[45]. In the 1990s, Density Functional Theory (DFT) allowed for the prediction of properties like band gaps and phonons[46]. However, these methods were computationally expensive, taking hours per structure on supercomputers, limiting the scale of discovery. Although databases, such as the Inorganic Crystal Structure Database (ICSD) and NIST alloy data emerged, they were fragmented and lacked integration, hindering large-scale materials discovery.

*1.4 Emergence of Materials Informatics*

The concept of materials informatics began to take shape in the late 20th century, marking a shift from traditional trial-and-error methods to a data-driven approach. The term "materials informatics" was coined by John R. Rodgers at the Materials Research Society Fall Meeting in 1999[47] to integrate information science, data processing, and systems engineering into materials research, focusing on data generation, management, and knowledge discovery[48].

With increasing computational power and the advent of high-throughput experimental techniques, such as combinatorial materials science, the field moved from basic data collection to extracting correlations between processing-structure-property relationships using statistical learning and data mining. This shift laid the groundwork for Integrated Computational Materials Engineering (ICME), fostering collaboration between computational and experimental methods[48].

In the early 21st century, soft computing and statistical learning methods began addressing nonlinear problems in materials science. Chen et al.[49] developed an artificial neural network (ANN) model to relate yarn properties, fabric parameters, and shear stiffness using data sensitivity criteria for input variable selection. Sundararaghavan and Zabaras[50] applied X-means clustering and dimensionality reduction techniques to optimize microstructure design for specific materials. Broderick et al.[51] utilized data mining to classify crystal structures based on electronic structure data, offering a novel approach that contrasted traditional methods.

By 2011, it was clear that traditional empirical methods were insufficient for addressing the growing demands of industries like aerospace, energy, and electronics. The limitations of previous approaches, combined with high experimental costs and slow computational simulations, highlighted the need for a more systematic, data-driven approach. MGI's integration of high-throughput computational techniques and machine learning transformed the landscape by accelerating material discovery and enabling predictive modeling of material properties in real-time. MGI funded large-scale databases, such as the Materials Project, and catalyzed the development of machine learning models capable of predicting material properties in seconds, dramatically reducing the cost and time required for material discovery[43].

## 2 Active Learning for Targeted Materials Exploration

The materials challenge, in its full generality, encompasses a remarkably high-dimensional discovery or search space, potentially containing millions of possible compounds. Of these, only a minuscule fraction has been experimentally explored. This vast space spans complex interdependencies across chemistry, crystal structure, processing conditions, and microstructure. Furthermore, compounds can be multicomponent (e.g., solid solutions), and their properties are often intricately dependent on materials descriptors or features operating at multiple length scales.

By about 2012, the state of art to tame the complexity by rational design as depicted in Figure 4 was to first collect all the chemistry and structural information relevant to a problem. Next, define the descriptor or feature space (the bits of information or genes) that could consist of bond length, bond angle, energetics from first principles calculations or from thermodynamics, and create a surrogate model for the property or objective of interest by using off the shelf python scripts from Scikit-learn. In order words, one typically performed supervised learning tasks on labelled data with various machine learning algorithms (e.g. SVM, SVR, KNN, NN) for classifiers and regressors to predict the objective. The model was then applied to a large space of potential or unexplored candidates, and the predictions were tested via first principles calculations on relatively simple systems where possible. It was rare for the predictions to be experimentally validated to find new materials. Concurrently, unsupervised learning techniques were applied to unlabeled data to glean insights into patterns or groupings within the dataset. This machine learning approach operated in parallel with the development of databases (e.g., Materials Project, OQMD, AFLOW) derived from electronic structure calculations, as well as high-throughput combinatorial experiments to successively down-select the vast search or candidate space[52].

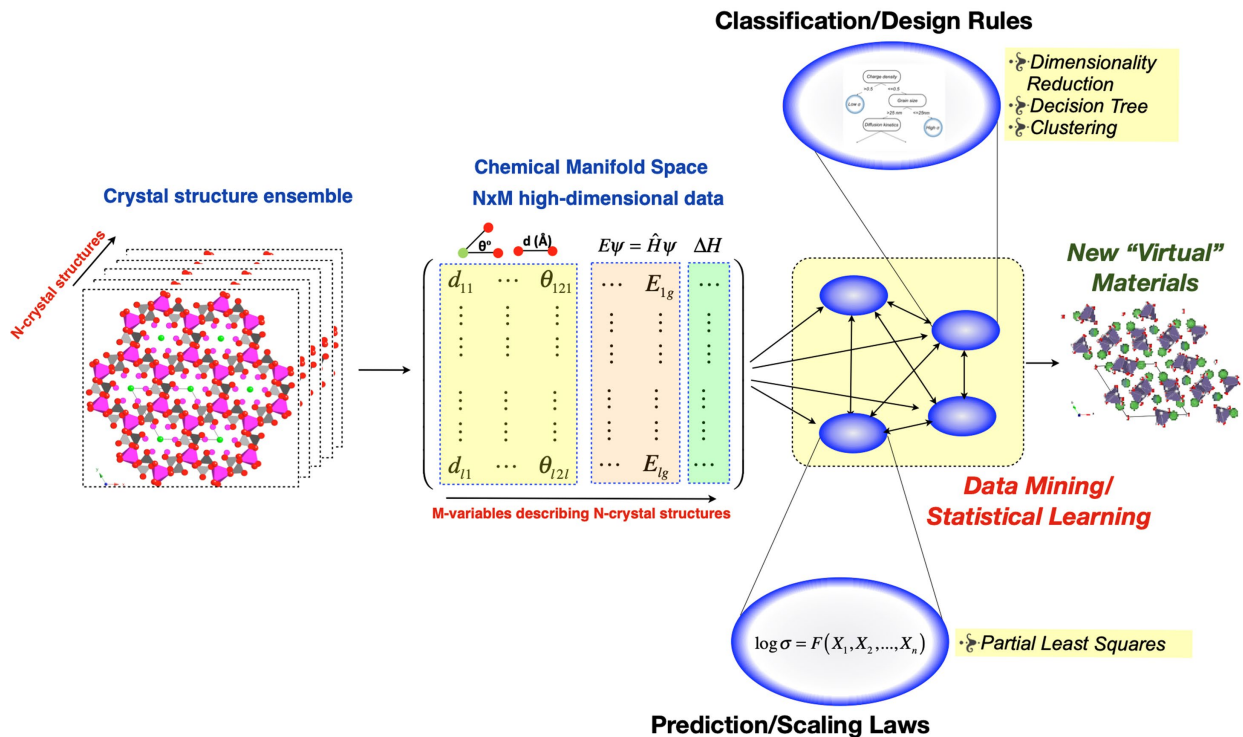

Figure 4. A serial approach to learning from data and making predictions using standard off the shelf ML approaches, as largely studied ~ 2012 – 2018.

It became apparent to some of us, then at Los Alamos National Laboratory, that for any given materials problem the overarching objective was to find a means of accelerating the process of judiciously selecting chemistry, composition, processing conditions leading to the requisite microstructure with the desired properties, thereby ensuring targeted performance. Clearly, this was a complex optimization, control and learning problem requiring to search a high-dimensional feature space. In addition, materials discovery in a laboratory environment involves relatively small amounts of data and inherent uncertainties in measurement. Moreover, material systems can be multicomponent, contain defects, and their feature search space can be inordinately large. Therefore, experimental observations are often difficult and time-consuming, necessitating a strategy to maximize the value of each observation. This underscores the need for design, which can be quantified by defining the *utility* of an experiment. This utility, $u$, needs to be maximized by some measure to select the next experiment to perform. That is, we define $x_t = argmax_x u(x|D_{1:t-1})$, where $u$ is the utility or acquisition function given the data, which is maximized in order to select at instant, $t$, the material, $x$ for which $u$ is a maximum. Thus, a decision theoretic approach can be used to recommend candidates for experiments in an adaptive manner.

This adaptive framework is the foundation of Active Learning (AL), a field whose historical trajectory has been shaped by a shift in purpose between Computer Science (CS) and Materials Science. While both fields utilize AL to guide selection, they are driven by two distinct motivations: statistical efficiency and exploratory discovery. While rooted in the foundational principles of Optimal Experimental Design from statistics, pioneered by Smith, Kiefer, and Fedorov to minimize the estimator variance in a regression problem to make the most use of experimental resources, its evolution reveals a change in emphasis[53]. In Computer Science, AL emerged in the late 1980s to address the "labeling bottleneck." The advent of "Query Learning" demonstrated that a learner could identify a target concept (e.g., a grammatical rule) exponentially faster by generating its own informative examples for an oracle to label, compared to learning from passive, random samples[54]. This evolved into the practical "pool-based" paradigm as the internet provided vast, unlabeled data reserves[55]. The task was no longer to synthesize queries, but to intelligently select the most *informative* instances from a large, static pool for human annotation, formalizing AL as a standard machine learning tool[56].

In domains such as text classification and computer vision, AL's objective was unequivocally model-centric. With unlabeled data abundant and labeling costly, strategies like uncertainty sampling and query-by-committee were developed to build the most accurate *global predictive model* with the minimal labeling budget. The success metric was overall test accuracy; the goal was to create a model that performed well *everywhere* in the feature space.

The application of AL in contemporary materials and chemistry informatics retains the algorithmic framework but pivots the objective towards space-centric exploration. The goal is no longer a universally accurate model, but rather an 'outlier hunt' which is the targeted discovery of optimal regions within a vast chemical or materials space. Acquisition functions shift from pure uncertainty towards expected improvement or knowledge gradient, aiming to iteratively guide expensive experiments or simulations (e.g., Density Functional Theory) toward materials with desired extreme properties.

This aligns well with the rise of autonomous experimentation. In a "self-driving lab," the AL agent acts as a closed-loop decision-maker: it proposes the next synthesis condition or composition, the system executes the experiment, and the result updates the model, creating a hypothesis-generating engine for physical discovery[8]. This transition thus creates a fundamental distinction: In classical ML, the model is the valuable end-product; data selection is merely a means to cost-effective training. In Materials Informatics, the physical material is the end-product; the model serves as a temporary, evolving guide to navigate the search space. This evolution has paved the way to transform AL from a data-efficiency tool into a primary driver of scientific discovery.

*2.1 Bayesian Optimization: Utility-Driven Experimentation*

The adaptive design strategy was initially proposed in 2014[57,58], but was not experimentally validated[18] until 2016 when ultra-low thermal hysteresis (1.84 K) NiTi-based shape memory alloys were found from a candidate space of approximately 800,000 compositions. Figure 5 shows the active learning loop and the key idea of iterative feedback using new data from experiments and/or computations that were previously recommended. Optimal experimental design leveraging uncertainties was critical in balancing the trade-off between exploiting the predictions of the surrogate model formed from data and exploring further the search space. A number of utility or acquisition functions can be defined, in terms of the mean and standard deviation of the predictions from the surrogate model to rank candidates so that the one that maximizes the function for suggesting further experiments. The results from experiments then augment the data set leading to an improvement in the model for the next iteration. The concept has been used widely, however, there are a number of issues with the approach, which we consider below.

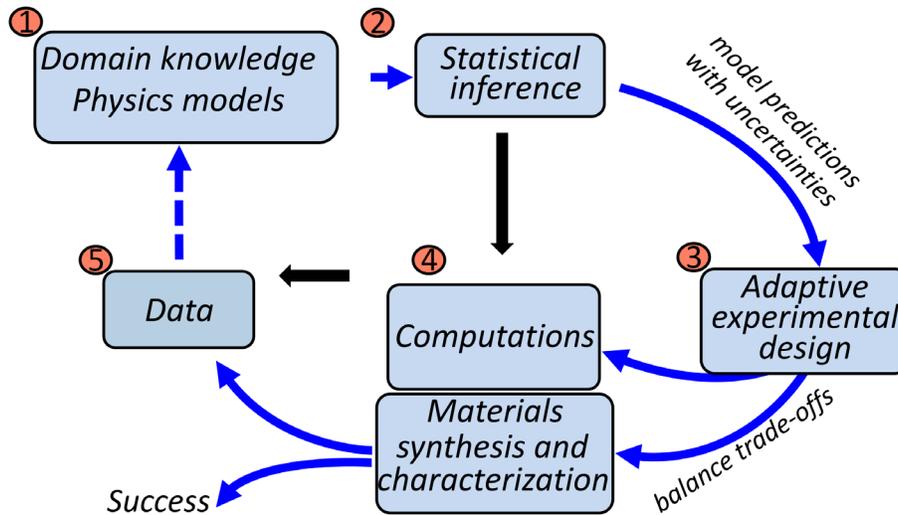

Figure 5 The adaptive design loop proposed in Ref[57], [58] and experimentally demonstrated in Ref[18]. Prior knowledge, including data from previous experiments and/or physical models, and relevant features serve as input to describe the materials. A surrogate model, such as a Gaussian process, makes predictions with error estimates. An experimental design tool suggests new experiments (synthesis and characterization) based on exploration/exploitation strategies with the dual goals of model improvement and materials discovery. The results augment a database for the next iteration of the design loop. Adapted with permission. [18] Copyright 2016, Springer Nature under a Creative Commons Attribution 4.0 International License (http://creativecommons.org/licenses/by/4.0/).

The strategy employed above, at the heart of experimental design, is an example of sequential decision that cuts across many fields of science and engineering as the same type of problem is being studied in different fields[59,60]. The question addressed is: what is the optimal way to make decisions? That is, what are the sequence of actions one needs to perform to obtain the best results? In computer science, the study of sequential learners and reinforcement learning is well known. In Engineering sciences, it is studied under optimal control. In the neuro sciences, understanding the brain and studies on the basis of human decision making are important areas of study. In psychology understanding how animals make decisions –again the reward system. In mathematics, studying the mathematics of optimal control (and applications to areas such as resource allocation and scheduling, the field known as Operations research). In economics, we have game theory and utility theory, that is, how do people make and rationalize decisions. Hence, sequential learning is of interest in many areas. It has been important in industry where data is limited and the challenge is to guide complex codes and or experiments which are inherently resource hungry. Thus, functions, such as probability of improvement and expected improvement using uncertainties were developed as aids to guide the next outcome by maximizing the expected utilities in an iterative Bayesian Global Optimization (BGO) loop[8,61].

The Bayesian approach has been used extensively on many different materials design problems involving manual synthesis and characterization in the laboratory. It has also been employed autonomously in self-driving laboratories for optimizing properties of thin-film materials by modifying the film composition and processing conditions to improve on the target response. Initial applications include optical and electronic properties of transport materials as well metallic films formed from combustion synthesis[15,16]. Since then, self-driving laboratories have and are been constructed for the study of ferroelectric ceramics as well as the synthesis of alloys using laser printing and mixing metallic powders. Recently, the autonomous synthesis of solid-state inorganic powders was demonstrated by using predictions from computational first principles to drive experimental protocols[62]. After 17 days of operation, the active learning loop synthesized 41 compounds from 58 initially targeted. The synthesis conditions were obtained by using a model trained on data from the literature using text mining and where the conditions needed modifications, an active learning loop is employed incorporating aspects of thermodynamics.

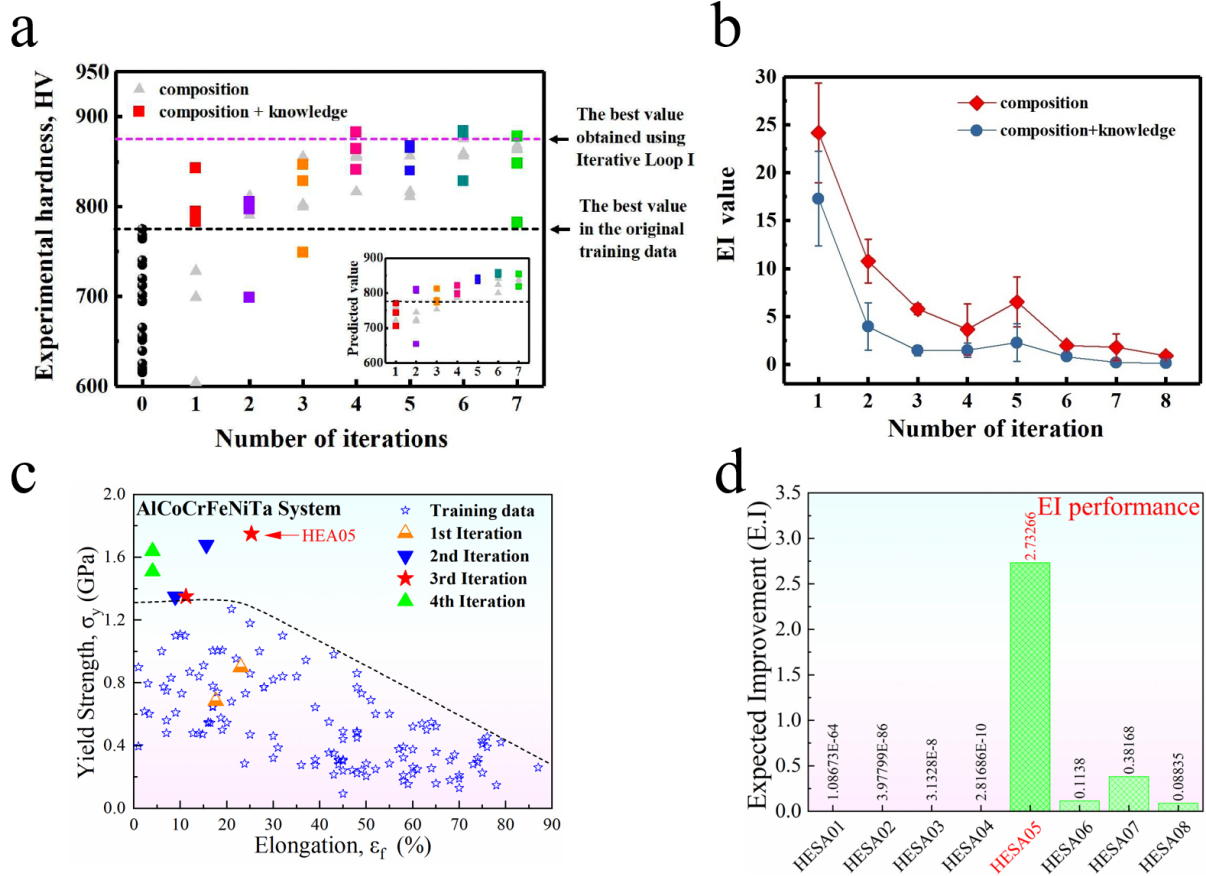

Figure 6. The issue of the lack of a clear stopping criterion that ensures no further improvement is likely illustrated for High Entropy Alloys (HEAs) showing the behavior of the expected improvement, EI $=\sigma[\varphi(z)+z\varphi(z)]$, where $\varphi(z)$ and $\varphi(z)$ are the standard normal density and cumulative distribution function, respectively as a function of experimental iterations. a) The hardness of new synthesized alloys as a function of iteration number. The black dotted-line represents the best hardness in the training dataset and the magenta line the best value obtained using composition only[24]. The inset are the predicted values showing a similar tendency to the measured values. The best performer is found on the 4 th iteration but EI in (b) peaks at about the 5 th iteration. c) A comparison of YS vs. elongation for HEA training data showing that synthesized alloys outperform those in the training data after the first iteration[63], however, after the 5 th iteration EI fluctuates without alloys with better performance. (a,b) Reproduced with permission. [24] Copyright 2019, Acta Materialia. (c,d) Reproduced with permission. [63] Copyright 2025, Springer Nature under a Creative Commons Attribution 4.0 International License(http://creativecommons.org/licenses/by/4.0/).

Although Bayesian Optimization has been widely employed in materials science to navigate complex design spaces, there are a number of issues with it. To begin with, the usual approach taken is single step or one step look ahead optimization balancing exploration and exploitation. Although two-step and multistep schemes have been proposed, they are highly cumbersome or impractical to implement even if they are meant to capture the interdependence of consecutive design decisions. As such, it can be adequate for relatively small search spaces, but its performance can significantly be affected as the dimensionality of the search space increases, as shown recently by Xian et al.[64] It can miss promising regions as it struggles to explore the space in high dimensions. However, a key issue with even the basic scheme is the absence of a clear stopping criterion which makes it difficult to determine when the optimization should terminate. There are numerous examples in the literature where an acquisition function, such as expected improvement, EI, is very small as the iterations proceed and then increases, often with findings that are desirable, and then decreases again.

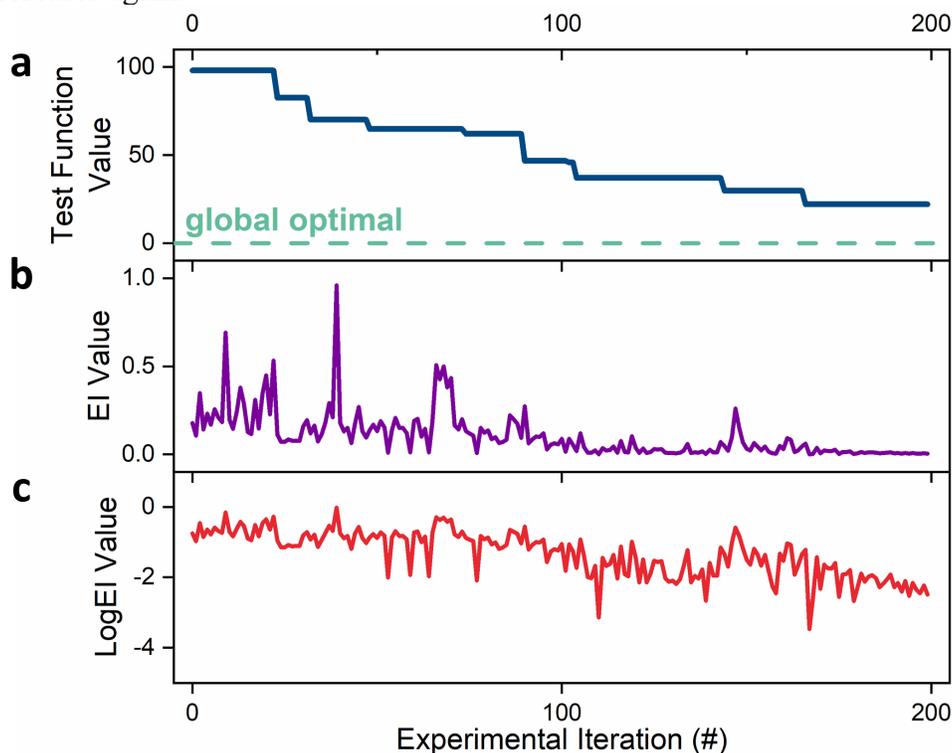

Figure 7 The fluctuations in the Expected Improvement values for the Rastrigin function a) Best-so-far values for the function itself, b) the fluctuations in EI values and c) the decreasing trend in logEI values as a function of experimental iterations[65]. Reproduced with permission. [65] Copyright 2025, Springer Nature under a Creative Commons Attribution 4.0 International License (http://creativecommons.org/licenses/by/4.0/).

Figure 6 shows a couple of examples of EI for materials design problems from the literature as a function of iteration number or number of times new experiments are carried out, with the corresponding value of the objective. Most studies terminate the iterative loop when the authors are satisfied with the results or run out of budget. However, the question remains whether a better solution would have been obtained were the iterations to continue. As pointed out by Dalton[66], as EI values often become very small or approach zero in later iterations, this leads to numerical instability that affects the gradient based optimization. This can be addressed by replacing EI with logarithmic Expected Improvement (logEI)[65]. Thus, if EI approaches zero from R+, logEI tends to negative infinity, ensuring greater numerical stability. Figure 7 shows the improved convergence in ten dimensions, d, for the synthetic Ackey function defined by $f(x) = -20 \cdot exp\left(-0.2 \cdot \sqrt{\frac{1}{d}\sum_{i=1}^{d} x_i^2}\right) - exp\left(\frac{1}{d} \cdot \sum_{i=1}^{d} cos(2\pi \cdot x_i)\right) - 20 + exp(1)$ with $f(x^*) = 0$, $x^* = (0, \ldots, 0)$, compared to EI as recently studied by Xian et al.[65]

## 2.2 Reinforcement Learning for High-Dimensional Materials Design

As the materials feature space increases, as typically occurs if more elements or components are added to expand the materials class, it is more challenging with BGO in the course of sequential design to get close to the target solution, even if continuous gradient optimization is employed instead of discrete space. To overcome this limitation of BGO in high-dimensional spaces, Xian et al.[67] began to explore the application of Reinforcement Learning (RL) to materials design as this bottleneck or combinatorial explosion is precisely what was overcome in 2015 by a deep neural network agent trained to play Atari games at a level comparable to a human player.

Using a deep Q-network and following a Markov decision process, an RL agent succeeded in playing a portfolio of 49 games by interacting with the "environment", receiving pixels and a score as reward, without any modifications to the algorithm. This success was followed by AlphaGo, which was the first to defeat a world champion at the game Go in which there are $10^{170}$ game positions in principle. Although AlphaGo employed supervised learning by utilizing human expert moves from set positions, AlphaGo Zero did not use any supervised learning and was purely the result of the agent interacting with the environment by performing actions to change its state and make decisions on the next action based on the resultant reward in changing the prior state[64,68,69]. The basic hypothesis is to maximize the cumulative reward, the sum of immediate and future rewards in the Markov process. Since then, RL has been successfully applied in various fields from playing adversarial games to controlling real-time robotics. It's been a mainstay in mimicking human sensing and decision making abilities so that an agent or a robot can interact with the environment it is in, interpret data, and taking actions to adapt. Recently, its use in LLMs, initially with human feedback in ChatGPT, but more recently upfront in DeepSeek to allow for more reasoning in solving problems in a step by step manner, has led to considerable improvement in the performance of LLMs.

This ability to navigate large candidate feature spaces and make decisions appropriately via certain actions to incorporate future outcomes offers unique advantages in sequential materials optimization and discovery. Given the large number of self-driving laboratories being built currently for materials design, it is necessary to compare the relative strengths of different sequential design strategies to understand how well they perform in sampling high value regions of the feature space. Recently, Xian et al.[67] demonstrated a proof of principle of how RL can be used for materials design by efficiently synthesizing compositions of NiTi based alloys with large enthalpies of transitions. A game-playing agent makes decisions on the composition of a new component and interacts with an iteratively improved surrogate model of the reward. It modifies the composition, if favorable. The objective is to maximize the cumulative reward, the sum of the immediate reward for the next action and discounted future rewards, assuming a Markov process. TiNi-based SMAs undergo a reversible solid to solid (SS) phase transformation and the objective is to learn compositions, such as Cu, Hf, Zr that can be added into the host TiNi alloy to maximize the enthalpy of transition. Figure 8 shows how the design starts from a single component, which is then substituted for another component of a given composition. The design chain continues with subsequent actions until all available components are considered. A reward is assigned based on the difference of the target before and after the action (old and new compositions). The RL agent employs a neural network to estimate the reward for every action of assigning compositions (the Q value). The reward is obtained from the surrogate model constructed from training data, which is iteratively improved via experiments suggested after the inner RL loop converges. The strategy based on RL leads to the phase change multicomponent alloy $Ti_{27.2}Ni_{47}Hf_{13.8}Zr_{12}$ with a transformation enthalpy ($\Delta H$) - 37.1 J/K, one of the highest values obtained in this class of alloys.

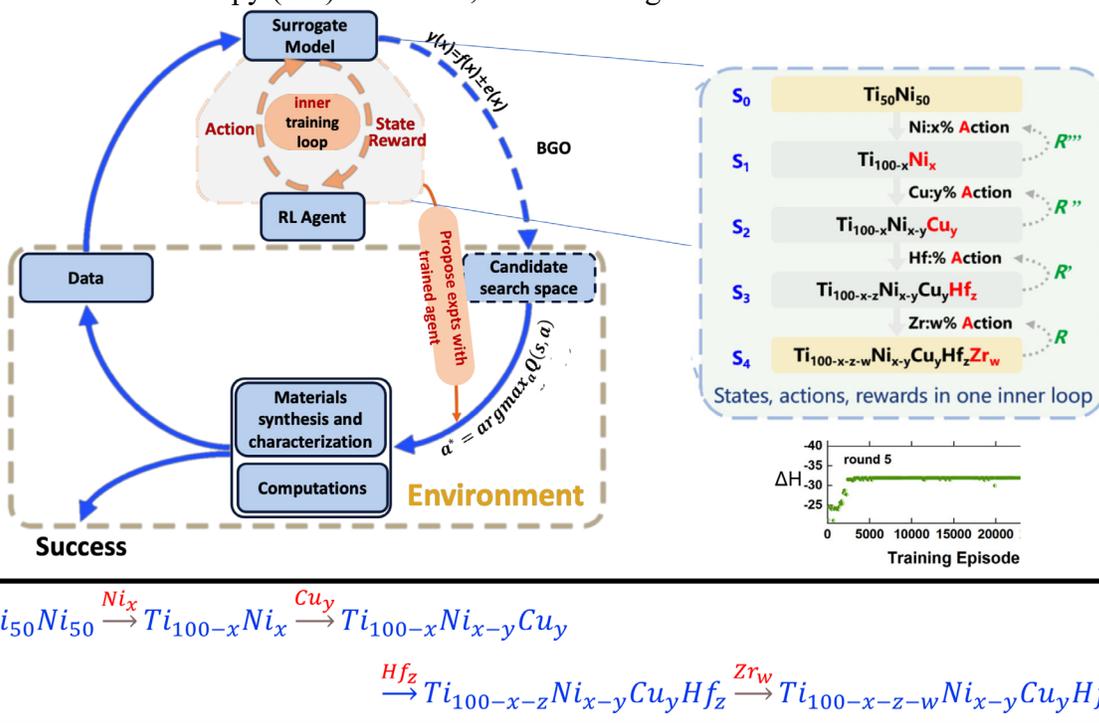

Figure 8 An overlay comparing Bayesian Global Optimization and model-based Reinforcement Learning for materials design. The RL agent executes the inner loop by choosing (taking action) a given element composition that defines a state and obtains a reward for it from a surrogate model. An episode consists of defining all the element compositions. The agent navigates the composition search space to maximize the cumulative reward and the final result (e.g. Q table) serves as a basis for choosing experiments. In discrete BGO, the candidates in the search space are ranked so that the utility or acquisition function is maximized. Adapted with permission. [67] Copyright 2024, Acta Materialia under a Creative Commons Attribution 4.0 International License (http://creativecommons.org/licenses/by/4.0/).

Given the limited depository of data for alloys, it is important that it can be utilized to speed up the learning process. Figure 9 shows how existing offline data can be employed in pretraining an alloy data set so that regions of higher values in the feature space can be accessed more frequently. With feedback from new experiments an agent can gradually explore new regions with higher values that are compositionally differently from the initial dataset (Figure 10).

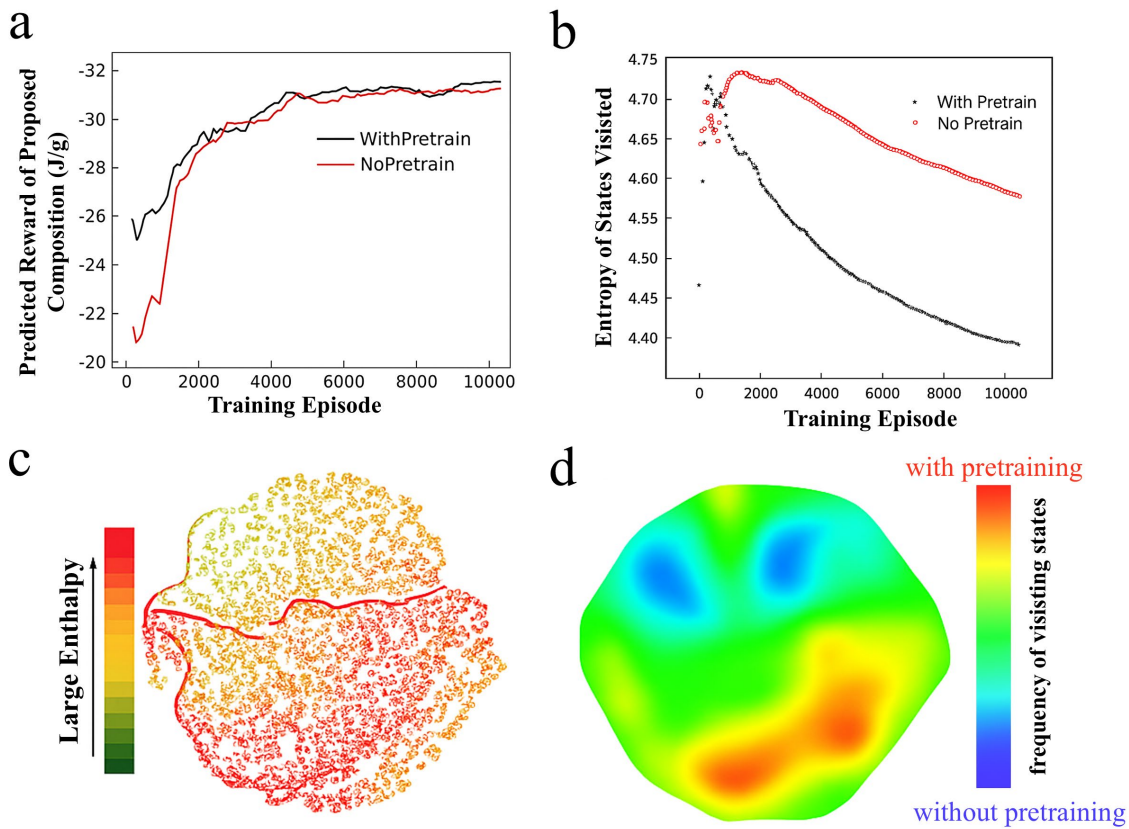

Figure 9 a) Effects of pretraining the RL agent with initial experimental data to minimize training episodes The order of component substitution of the alloy is not predetermined. Rolling-averaged predicted enthalpy of proposed compositions during the training process, b) State entropy (uncertainty) based on predicted enthalpy histogram during the training process. Agent shows greater frequency of visits in the high enthalpy value regions for the pretrained case compared to without pretraining. c) t-SNE plot for all visited states (compositions) where the color represents the corresponding predicted enthalpy and d) frequency of compositional states visited. Shown is the difference between pretraining and no pretraining[67]. (a-d) Reproduced with permission. [67] Copyright 2024, Acta Materialia under a Creative Commons Attribution 4.0 International License (http://creativecommons.org/licenses/by/4.0/).

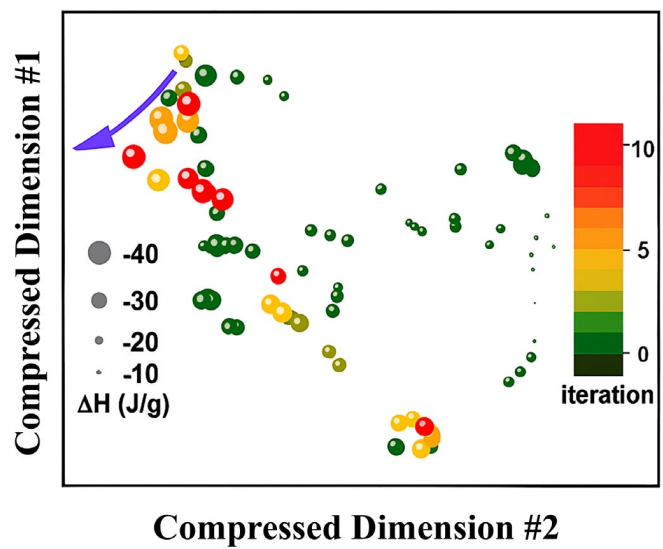

Figure 10 t-SNE plot showing all experimentally obtained compositions containing Ti, Ni, Cu, Hf and Zr[67]. The green compositions are the training data with the size of the circles representing enthalpy values. As the experimental iterations proceed, newer compositions drift further away from the green training data. Reproduced with permission. [67] Copyright 2024, Acta Materialia under a Creative Commons Attribution 4.0 International License (http://creativecommons.org/licenses/by/4.0/).

Finally, it is interesting to compare the relative merits of BGO-EI and RL in searching for high value regions in feature space. Ideally, this should be done within the context of a self-driving autonomous loop, but in its absence, we can compare their performance on a synthetic function which can capture to a degree the complexity of an experimental system. Figure 11 shows the search patterns for the 10-dimensional Rastrigin function using t-SNE dimensionality reduction. The scatter plots (Figure 11 c, d) show the sampled points for experimental iterations (#30-#150), with colors showing their proximity to the global optimum (red being close). The corresponding density estimate maps (Figure 11e) show the relative sampling preferences of both strategies, with yellow and blue regions corresponding to more frequently explored by model-based RL and BGO-EI, respectively. The visualization shows that BGO-EI tends to concentrate its sampling in more centralized regions, whereas model-based RL gives a more dispersed pattern. Both methods identify different high-value, though model-based RL appears to maintain better performance in finding near-optimal solutions.

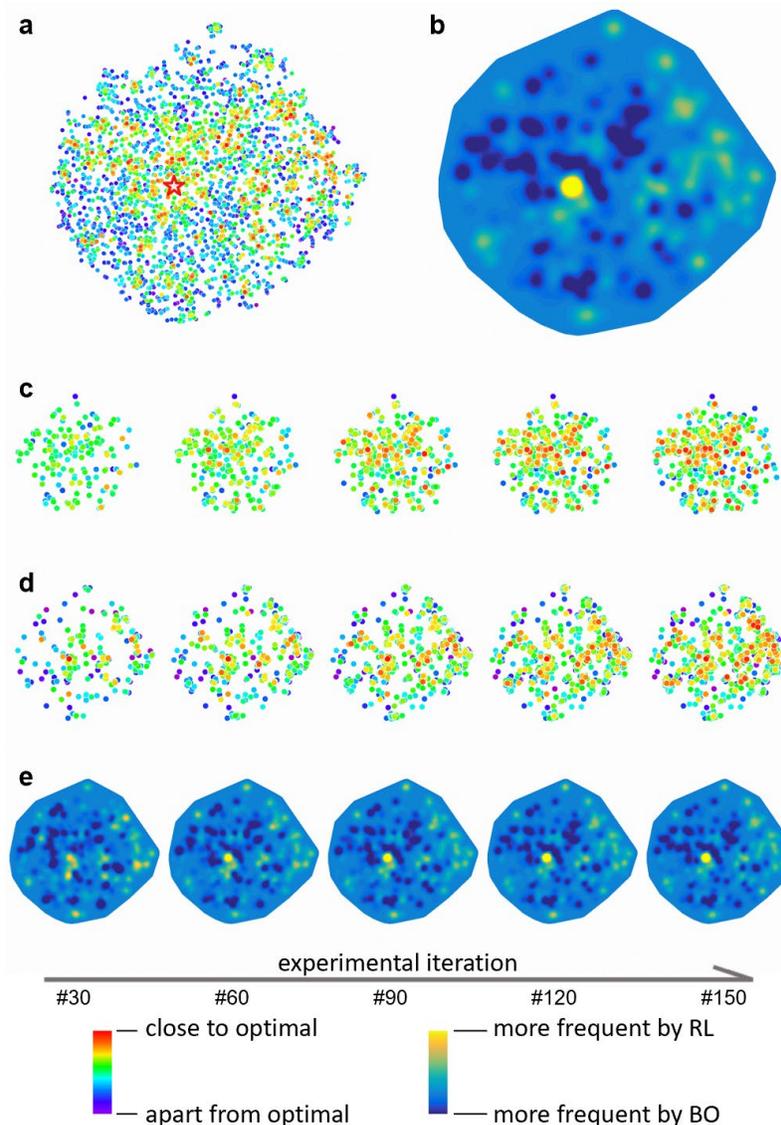

Figure 11 The sampling behavior of the RL agent as a function of iterations shown in visualizations of search patterns for the 10-dimensional Rastrigin function space using t-SNE dimensionality reduction. a) Sampling distribution with optimal point marked in red. b) Map of density of sampling showing density is highest at the global minimum. c), d) Time evolution of sampling points for BGO-EI and model-based RL, respectively. e) Corresponding density maps showing the relative exploration preferences of both methods with more dispersed sampling with RL relative to BGO-EI preferring regions in the annulus around the optimal [65]. Reproduced with permission. [65] Copyright 2025, Springer Nature under a Creative Commons Attribution 4.0 International License (http://creativecommons.org/licenses/by/4.0/).

*2.3 The Rise of Self-Driving Labs*

In recent years, the self-driving laboratory, also known as the "AI autonomous laboratory" has emerged as a key driver for accelerating materials discovery. This paradigm integrates robotics, artificial intelligence, and high-throughput experimentation to efficiently explore the vast material parameter space through a closed-loop feedback system guided by intelligent algorithms. Early pioneering work, such as the single-objective optimization of thin-film materials by MacLeod et al.[15] using the stationary "Ada" platform, and the development of a mobile robotic chemist by Burger et al.[70] capable of navigating standard laboratory environments, successfully validated the feasibility and immense potential of this concept. As the technology has matured, the research focus has expanded from optimizing single attributes to addressing the more complex trade-offs inherent in practical applications. A subsequent study by MacLeod et al.[16] serves as a prime example, employing multi-objective optimization algorithms to clearly delineate the Pareto front between different material properties. This provided an intuitive basis for decision-making to meet customized, multi-scenario demands and successfully translated laboratory discoveries into scalable processes.

A number of tools have been developed to learn from the results of each experiment, optimize parameters and make informed decisions for subsequent trials. These include MatterGen[19], which designs and suggests materials based on desired property prompts and MatterSim[71] that filters MatterGen's suggestions to predict stable candidates. Robotic arms and automated systems typically handle weighing powders, mixing solutions, and transferring samples between instruments. For example, RoboChem[72] performs a variety of reactions while producing minimal waste with a robotic needle collecting and mixing small amounts of reagents, which then flow through to a reactor via a tube. The reactor then uses LEDs for photosynthesis by triggering molecular conversion then analyzed by an NMR spectrometer which then feeds the results to the controller to guide the next steps of the experiment. Other platforms include Polybot[73], developed at Argonne National Laboratory for polymer synthesis, and A-Lab for the synthesis and characterization of inorganic materials from powders. The closed loop paradigm typically takes data from material databases, relevant literature, previous successful and failed experiments to propose candidates, which are synthesized by instructing robotic arms to mix precursors and heat and anneal them in a furnace. The characterization can be done by XRD to detect secondary phases and/or by electron microscopy to gauge material quality. After interpreting the data, parameters are modified to establish, for example, variations on the compositions or synthesis protocols and the cycle is repeated to iteratively refine the outcomes.

How successful have some of these initiatives so far been? RoboChem[72] can optimize the synthesis of 10–20 molecules in a week, a task requiring several months for a student to perform. The A-Lab[62] successfully synthesized forty odd inorganic compounds in just 17 days and exemplifies the shift in integrating a first-principles calculation database with synthesis knowledge extracted from literature. Pushing this paradigm to new heights, the research by Jiang et al.[74] features an "AI chemist" designed to tackle the ultimate challenge of in-situ oxygen production on Mars. By tightly coupling first-principles calculations with robotic experiments, it autonomously screened millions of potential formulations to identify an optimal catalyst. The trend is to refine these systems to yield high-quality, reproducible results and minimize human error and bias, and apply them to design next-generation batteries, catalysts for clean energy, and materials for extreme environments. Moreover, the tools are not only advancing synthesis optimization, but also deepening fundamental materials knowledge such as thermodynamic properties[75]. In addition, the multiscale integration of computational modeling with physical experiments, including imaging, and informatics creates a powerful feedback loop for autonomous discovery, data fusion and scientific understanding[76]. Thus, platforms, such as the Autonomous MAterials Search

Engine (AMASE)[77], can self-navigate the mapping of a phase diagram by continuously comparing theoretical and experimental data.

## 3 AI Foundations for Materials Science

The field of Artificial Intelligence (AI) encompasses any application in which a task is performed without human intervention, such as in self-driving cars or recommendation systems. Within AI, Machine Learning (ML) employs tools such as statistical inference on assembled data to perform prediction, forecasting, visualization, feature engineering, and analysis[78]. Deep Learning (DL), a prominent subtopic within ML, arose from attempts to mimic the human brain's complexity via neural networks layers of interconnected nodes (perceptrons) that transform inputs into a collective output.

The development of DL is inextricably linked to a series of architectural and algorithmic breakthroughs. The foundational backpropagation algorithm, introduced by Rumelhart et al.[79] in 1986, employed gradient descent and the chain rule to enable efficient weight adjustment in multi-layer networks. Subsequently, specialized neural architectures were developed: Convolutional Neural Networks (CNNs), notably applied by Yann LeCun[80] to document recognition in 1998 for image processing; Recurrent Neural Networks (RNNs) for sequence-to-sequence mapping; and Long Short-Term Memory (LSTM) networks to better handle long-range temporal dependencies where RNNs struggled with vanishing gradients. However, the training of deep networks remained challenging due to gradient instability. The widespread adoption in 2011 of the Rectified Linear Unit (ReLU) activation function, which significantly alleviated the gradient vanishing phenomenon[81], was an important advance. This was empirically validated by the success of AlexNet in 2012, which demonstrated the necessity of GPU acceleration and model depth for state-of-the-art visual recognition, comprehensively ushering in a new era[82]. More recently, the introduction of self-attention in transformers has revolutionized fields such as natural language processing[83].

In materials science, DL has triggered a paradigm shift by providing tools for its unique data challenges. A core initial difficulty was that molecules and crystals represent non-Euclidean data, unlike the regular grid of images. Effectively characterizing their micro-geometric and topological structures was once a central challenge. A major step was taken in 2015, when Duvenaud et al.[84] proposed Neural Graph Fingerprints. By applying differentiable convolution operations on molecular graphs to replace traditional fixed fingerprints, they achieved end-to-end molecular feature learning, significantly improving prediction accuracy for properties like drug solubility. To address the unique infinite periodic structures of inorganic materials, Xie and Grossman[85] developed the Crystal Graph Convolutional Neural Network (CGCNN) in 2018. This framework constructs a crystal graph to encode atomic information and bonding interactions, overcoming limitations in handling arbitrary system sizes and achieving prediction accuracies comparable to Density Functional Theory (DFT) for properties such as formation energy and band gap

Beyond property prediction (the forward problem), DL has propelled the inverse design of materials. For molecules, Gómez-Bombarelli et al.[86] showed how the use of Variational

Autoencoders (VAEs) to map discrete molecular representations (SMILES strings) into a continuous latent space, enabled gradient-based optimization to search for novel molecules with specific properties. For the more challenging generation of periodic materials, generative AI has moved beyond traditional evolutionary algorithms. As highlighted by Madika et al.[87], there has been rapid progress in crystal generative and diffusion-based approaches for diverse materials datasets[88], dramatically advancing crystal structure prediction and microstructure design. Platforms range from CNN-based frameworks, e.g. Crystalyze[89] to advanced diffusion models. Notably, Xie et al.[90] integrated Diffusion Models with VAEs to propose CDVAE, incorporating noise-conditional score networks and Langevin dynamics to resolve difficulties with periodic boundary conditions and invariance, achieving a breakthrough in generating realistic and diverse crystal structures. Other state-of-the-art diffusion-driven models include MatterGen[19].

Deep Learning Molecular Dynamics (DLMD) stands out as a transformative application, directly addressing the long-standing "accuracy-versus-efficiency dilemma" in molecular simulations. Machine Learning Potentials (MLPs) use DL frameworks to learn a representation of the many-body potential energy surface from high-accuracy quantum-mechanical data (e.g., DFT), bridging the gap between computationally expensive ab initio methods and less accurate empirical force fields. As a result, MLPs achieve near-ab initio accuracy while maintaining computational efficiency competitive with empirical potentials. Prominent open-source software packages enabling this include DeePMD-kit[91,92] and GPUMD[93,94], which allow for highly efficient, large-scale atomistic simulations previously considered intractable.

Furthermore, the emergence of Materials Knowledge Graphs (MKGs) is impacting autonomous materials discovery. This involves using machine learning and Natural Language Processing (NLP) methods to automatically extract and structure materials data from the scientific literature. Venugopal et al. demonstrated this feasibility in building MatKG[95]. Subsequent systems, like the LLM-based MKG introduced by Ye et al.[96]., integrate multidisciplinary materials data to enable predictive links between materials and their applications More recently, Bai et al.[97] extended this paradigm to framework materials with KG-FM, achieving accurate information retrieval and reasoning by coupling large language models with graph databases. These studies collectively mark a transition from static materials databases to dynamic, AI-driven knowledge systems capable of learning, reasoning, and accelerating discovery.

*3.1 Understanding Generative vs. Discriminative AI Models*

In the context of AI and machine learning (ML), a foundational distinction exists between *discriminative* and *generative* models. Discriminative models focus on learning the boundary between different classes of data. They estimate the conditional probability distribution, $P(y\,|\,x)$, which represents the likelihood of an output label *y* given an input *x*. These models are commonly used for tasks like classification, regression, and structured prediction. Examples include support vector machines (SVMs), logistic regression, and many deep learning architectures when trained on labeled datasets.

In contrast, generative models aim to learn the joint probability distribution, $P(x, y)$, enabling them not only to classify data but also to generate new instances. By modeling how the data is generated, they can simulate new samples that resemble the training data. Examples of generative models include Variational Autoencoders (VAEs), Generative Adversarial Networks (GANs), and autoregressive language models like GPT. For example, GANs consist of two components: a *generator*, which creates new data samples, and a *discriminator*, which tries to distinguish real data from generated data. The adversarial interplay between the two networks leads to increasingly realistic data generation.

Transformers, especially in the form of large language models (LLMs), fall under the category of *generative* deep learning models. They use attention mechanisms and are trained to predict the next word (or token) given a context, effectively learning a probability distribution over sequences. However, these models can also be fine-tuned for discriminative tasks (e.g., classification, regression) by adapting the architecture and training objective.

In summary, the distinction lies in whether a model *predicts labels given inputs* (discriminative) or *models how the data itself is generated* (generative). This difference has important implications for applications in materials science: (1) Discriminative models are well-suited for property prediction, classification of phases, or defect detection; (2) Generative models enable inverse design, text-to-property inference, and the creation of new candidate materials, especially when data is sparse.

*3.2 NLP and Transformers: Tools Reshaping Materials Science*

Large Language Models (LLMs) are a class of neural network architectures fundamentally built upon attention mechanisms, extensively pretrained on massive datasets. Natural Language Processing (NLP), powered by advances in machine learning (ML), has profoundly transformed various aspects of daily life, from commerce and social media interfaces to speech recognition and machine translation. Given this widespread impact, it was a natural progression to investigate these powerful NLP tools for applications in the chemical, biological, inorganic, and broader materials science domains. Initially, these tools were instrumental in building extensive databases of reactions and biological pathways from existing textual data. Materials science possesses a vast and rich corpora of unstructured published and unpublished text detailing materials, their properties, and diverse synthesis and processing routes. The goal therefore is to first build databases of structured data and extract insights from this information, in as an automated manner as possible, and to train the transformers so that new materials with desired properties may be synthesized[98–102].

The dramatic advances in Large Language Models (LLMs) and Natural Language Processing (NLP) have been driven by pivotal concepts such as word embeddings and self-attention. These techniques let us decipher a word's linguistic meaning and its semantic interconnections with other words within a given context. Initially, word embeddings offered a "static" representation where words with similar meanings had similar numerical representations. But these early embeddings didn't inherently encode the ordering or contextual nuances of words within a sentence or sequence.

The introduction of the self-attention mechanism revolutionized this by creating a "dynamic" representation[103]. This mechanism allows a model to weigh the importance of different words in an input sequence when processing a specific word. By focusing on the most relevant parts of a word or phrases in a sentence, self-attention effectively encodes contextual information. This lets the model predict a target word based on the intricate context of its surrounding words, which is crucial for understanding the complex relationships within materials science literature. Thus, many of the initial LLMs for materials science leveraged pretrained models (e.g. BERT), with text input for materials features, such as compositions and processing.

*3.3 Mining the Literature: From Text to Data*

One of the first applications of text mining to inorganic systems was by Kim et al.[2] , who leveraged neural networks and parse-based methods to identify and extract synthesis parameters for metal oxides from over 640,000 journal articles. This involved recognizing synthesis verbs (e.g., sinter, dissolve, mill) and then, using dependency trees, extracting associated numerical values like sintering temperatures and stirring speeds. The authors analyzed the distributions of these parameters and subsequently employed machine learning to predict synthesis parameters for related systems, such as titania nanotubes.

Focusing on materials properties, Court et al.[104] built a database of 39,822 records, specifically extracting Curie and Néel phase transition temperatures from a corpus of 68,000 journal articles. More recently, in the realm of materials discovery via text mining, Wang et al. initially extracted chemical composition and property data for superalloys from 14,425 journal articles. Recognizing that supervised deep learning methods for Named Entity Recognition (NER) or relation extraction typically demand large, hand-labeled datasets—which might be scarce for a relatively smaller corpora like theirs—they opted for a rule-based NER approach combined with a heuristic text multiple-relation extraction algorithm[105].

Building on the utility of extracted databases, Wang et al.[106] further demonstrated the power of text mining for materials discovery. They trained a machine learning model to predict the $\gamma'$ Solvus temperatures of 15 superalloys that were reported more recently and thus not included in their initial extracted dataset. The model's predictions achieved a remarkable mean relative error of 2.27%, a level of accuracy comparable to data typically assembled from in-house experimental measurements. Even more significantly, this approach enabled the prediction and subsequent experimental synthesis of three previously unexplored Co-based superalloys: Co-36Ni-12Al-2Ti-1W-4Ta-4Cr, Co-36Ni-12Al-2Ti-1W-4Ta-6Cr, and Co-12Al-4.5Ta-35Ni-2Ti, all of which were predicted to have a $\gamma'$ solvus temperature exceeding 1250 °C. The experimentally measured temperatures for these novel alloys showed excellent agreement with the predictions, exhibiting a mean relative error of just 0.81%. This successful validation highlights the direct pathway from text-mined data to the targeted discovery and experimental realization of new materials.

*3.4 Specialized LLMs for Materials Prediction and Design*

Transformer models for different materials systems are numerous in materials science. From PolyBERT[107] and TransPolymer[108] to BatteryBERT[109], MatSciBERT[110], OpticalBERT[111], AlloyBERT[112] and SteelBERT[100], these are pretrained on the subject matter data and the input to these are aspects or attributes of the materials system specifying a given sample. For example, for alloys it would be composition, processing and other features such as temperature or even microscopic parameters of the material. Applications of LLMs typically include an end-to-end pipeline for multiple prediction tasks. For example, PolyBERT has been pretrained on 100M polymer strings and can output electronic, optical, thermodynamic and mechanical properties of polymers with downstream predictors. The downstream tasks via a multi-perceptron are typically trained using a relatively small data set. The pretrained models can also be fine-tuned to make predictions for a different data set. LORA (low rank Adaptation) is one such approach to fine tuning. It essentially tracks the changes in parameter weights from those of the pretrained model by using lower rank matrices to reduce the number of parameters by matrix decomposition. Systematic and careful finetuning of several control parameters is crucially important if the LLM is to be sensitive to other data that deviates from that used in the pretrained model. Figure 12 shows the workflow of SteelBERT, that potentially can be used for designing new steels[100]. The LLM, with 0.18 billion parameters, is pretrained on a corpus of 4.2 million abstracts in materials science and 55,000 full-text articles on steels. Its architecture incorporates a disentangled attention mechanism for enhanced performance. After tokenization, the training corpus is fed into the DeBERTa model with 12 encoders consisting of 12 attention heads. A final dense layer with a softmax activation function is used to identify masked tokens, a common pretraining task.

SteelBERT generates embeddings with 768 dimensions for textual processing routes and chemical compositions to predict mechanical properties from natural language input. These embeddings then serve as input into a hybrid deep neural network model, which processes them through shared and specific feature layers to predict the yield strength, ultimate tensile strength, and total elongation. SteelBERT with frozen weights and the hybrid deep neural network model are trained together on a relatively small experimental data set. Subsequent fine-tuning on a specific dataset, for example, from refined laboratory experiments, can allow for better performance as indicated in[100]. The model has successfully predicted common mechanical properties of 18 new steels reported in 2022 and 2023, data which was explicitly excluded from its pretraining corpus. It achieved impressive coefficients of determination ($R^2$) values of 78.17% (±3.40%), 82.56% (±1.96%), and 81.44% (±2.98%) for yield strength (YS), ultimate tensile strength (UTS), and elongation (EL), respectively. Furthermore, SteelBERT leverages a laboratory dataset to create optimized processing protocols for austenitic stainless-steel by fine-tuning. With only 64 experimental samples of austenitic stainless steels (ASSs), it effectively searches a large candidate space and optimizes the best text sequence of fabrication processes, specifically by incorporating a secondary round of cold rolling and tempering. The resulting steel achieved a YS of 960 MPa, UTS of 1138 MPa and EL of 32.5%, representing the best performance reported so far among all 15Cr ASSs.

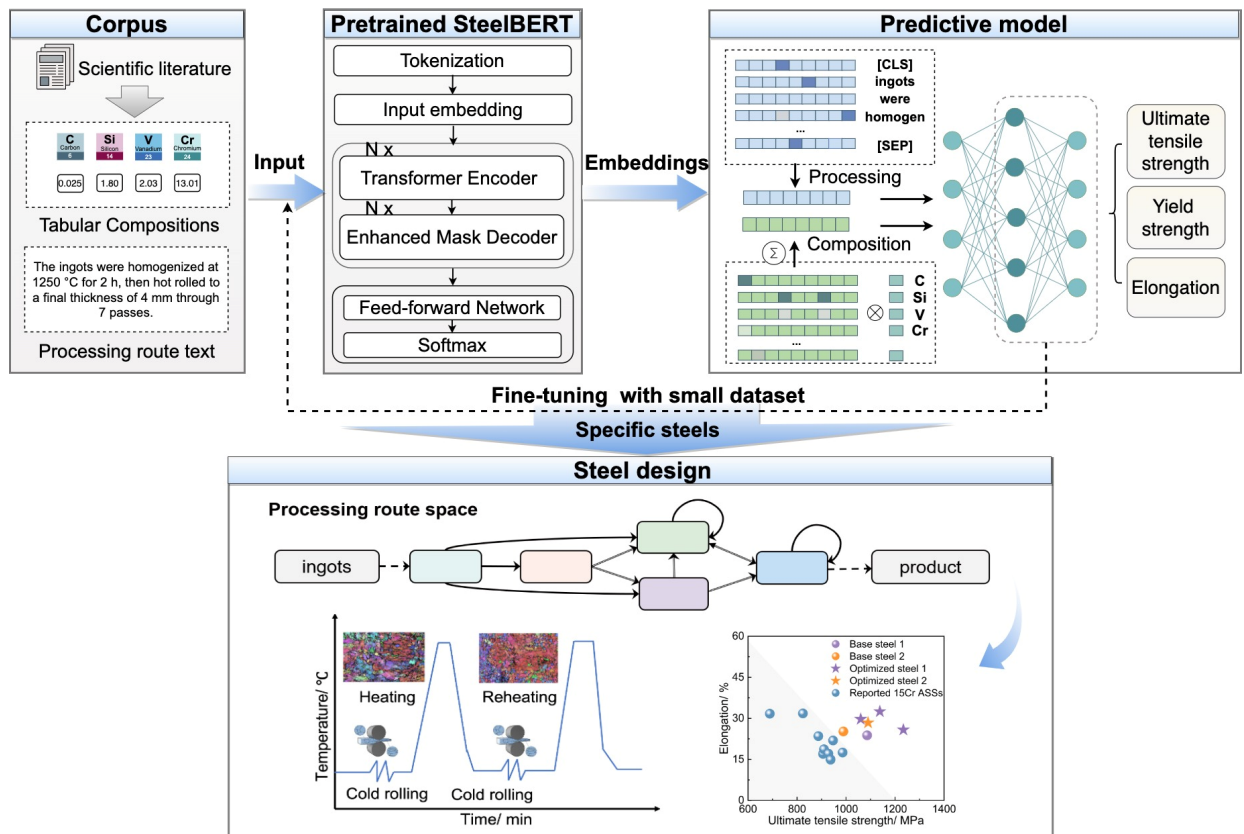

Figure 12 The workflow for SteelBERT encompasses data preparation, pretraining SteelBERT, making predictions with a multilevel perceptron using output embeddings from SteelBERT and fine-tuning for specific steels with small datasets[100]. Reproduced with permission. [100] Copyright 2025, Acta Materialia.

The success of predictive models clearly highlights their dependence on large, high-quality databases, but their manual creation remains a significant bottleneck. Addressing this challenge, the SLM-MATRIX framework by Li et al[113]. offers a powerful solution for automated knowledge extraction from scientific literature. Its innovation lies not in a single large model, but in a collaborative architecture of smaller, open-source models. The framework integrates two complementary reasoning paths: a Mixture-of-Agents (MoA) path that leverages the collective intelligence of diverse models to generate a broad set of candidate answers, and a parallel generator-discriminator path enhanced by Monte Carlo Tree Search (MCTS) that explores deep reasoning trajectories and validates their logical consistency. Crucially, only answers mutually verified by both independent paths are accepted, ensuring high fidelity. This robust methodology yields impressive accuracy (up to 92.85% on Bulk Modulus dataset[98]) at a fraction of the cost of proprietary models and without any fine-tuning. Ultimately, frameworks like SLM-MATRIX are essential for building the rich, structured datasets that power predictive models, thereby accelerating the entire data-driven materials discovery cycle.

To address the challenge of predicting complex properties like ductility, which is heavily influenced by processing and microstructural details often described only in text, Peng et al[114]. proposed an information fusion strategy. Their framework integrates a pre-trained language model, MatSciBERT[110], with contrastive learning to mine and utilize implicit knowledge from the scientific literature. The model demonstrated superior performance, achieving a coefficient of determination ($R^2$) of 0.849 on a titanium alloy validation set and 0.680 on a refractory multi-principal-element alloy (RMPEA) test set. Using this model, they performed ductility predictions across the Ti-V-Ta ternary compositional space without prior microstructural knowledge, identifying the $Ti_{30}V_{40}Ta_{30}$ alloy as a candidate with superior ductility. Subsequent synthesis and experimental testing validated the prediction, with the measured tensile elongation of 24% for $Ti_{30}V_{40}Ta_{30}$ closely matching the predicted value of 23.46%. This work illustrates how combining language models with materials data can accelerate the discovery of materials with desirable properties, even when explicit microstructural data is scarce.

However, multimodal models for different tasks or properties that depend on the literature and performance data, as well as those that integrate different modalities, such as microstructure and performance data, have also been recently introduced. The former, fall within the domain of Multi-task learning (MTL) integrating diverse information of related tasks or properties, and the latter within the scope multimodal learning (MML). In MTL, knowledge is shared between tasks to significantly improve the prediction accuracy of individual tasks to reduce overfitting[115–117]. In biomedical and organic chemistry and polymers, where tens of thousands of materials and hundreds of properties are involved, multi-task deep neural networks have been employed to analyze and predict various properties using datasets containing tens of thousands of materials and hundreds of properties[118]. This allows MTL to capture correlations within tasks to better predict than single-task models. Nevertheless, the sharing needs to be such that a group of tasks would benefit from training together as MTL is susceptible to noisy data and outlier tasks[119–121]. For alloy design, for example, using MTL is not straight forward as the datasets are typically smaller, in terms of possible material properties and size of data sets for each property[122]. Also, the relationships between the tasks can be quite complex and interdependent so that combining tasks can lead to a decline in overall performance[120,123]. Wang et al.[105] applied this approach to search for superalloys with several objectives including low density, suitable freezing ranges and processing windows and high $T_{\gamma'\text{-solvus}}$ temperature without detrimental phases and strong oxidation resistance. In contrast to training all tasks indiscriminately, they identified task groups through a scoring approach that combines average performance gain and positive frequency. Their encoder – decoder framework reduces average normalized error by 37.5 % compared to using a single task.

After its success in domains such as NLP and computer vision[124], several studies have extended MML to materials science by integrating diverse materials data obtained through multiple characterization techniques[125–129] The aim of these approaches is to increase a model's understanding of the material system and mitigate data scarcity to improve performance. However, these MML approaches remain limited by incomplete datasets due to experimental constraints and the high cost of acquiring measurements. For example, synthesis parameters are often readily available, whereas microstructural data, such as those obtained from SEM or XRD, are not only expensive but difficult to obtain. Thus, the performance of MML models deteriorates when modalities are missing[130] Moreover, existing methods lack efficient cross-modal alignment and do not provide adequate mapping mechanisms. In a case study on NCM cathode materials, Liow et al.[131] achieved a prediction error of less than 10% relative to experimental results, providing initial confirmation of the approach's reliability. In the MultiMat framework, Moro et al.[132], incorporate modalities with direct physical significance: crystal structure, Density of States (DOS), charge density, and machine-generated textual descriptions. It not only set new benchmarks in predicting properties, such as band gap and bulk modulus, but reduced errors by up to 10% compared to previous models.

Despite these advances, general-purpose multimodal models such as GPT and Gemini still face fundamental limitations in performing core scientific tasks that require deep understanding and complex reasoning. This gap was recently highlighted by Alampara et al.[133] through MaCBench, designed to test the models. They found that the models excel at basic perceptual tasks, such as identifying equipment, but exhibit deficiencies in spatial reasoning, cross-modal information integration, and multi-step logical reasoning, indicating that they rely more on pattern matching than on scientific understanding.

*3.5 Generalist vs. Specialist: Evaluating LLM Scope in Materials Applications*

It is interesting to question if it is necessary to have separate LLMs for each metallic family of alloys. After all, developing a separate pretrained model for each alloy system is time consuming and resource intensive. A base model for alloys, for example, that has been trained on a substantial amount of literature, should capture at the very least the salient features of related alloys faithfully, especially after some fine-tuning using data sets for particular alloys. Steels are multicomponent alloys with up to ten to twelve components, some of which are very small, and even though SteelBERT is a relatively small LLM with 0.18 billion parameters, we speculate that it should act as a reasonable model for alloys such as Al, Ti, HEA, Ni, Nb. Moreover, a follow up question is how well do general purpose LLMs, such as those in the Llama, Qwen, Gemini or DeepSeek series perform with materials input even though they are not particularly trained on any specialized materials literature. For example, Qwen3 with 14 billion parameters is trained on a range of high-quality data, including coding, STEM (Science Technology Engineering Medicine), reasoning, book, multilingual, and synthetic data using 36 trillion tokens across 119 languages. Figure 13 shows the predictions for alloys from LLMs on test data we recently made[134] for the ultimate threshold stress using recently published data. The data set consists of 831 7xxx Al alloys with given compositions and processing conditions with measured values for UTS[135]. The results from SteelBERT are also shown in Figure 4 after generating embeddings from it to input into a multi-layer perceptron (MLP) model—comprising an input layer, six hidden layers, and an output layer to predict properties.

All the data from the aluminum alloy training set were used to enable the LLMs in Figure 13 to learn the internal relationships within the data through few-shot prompting. The prediction results reveal significant performance discrepancies between the models. The performance of the domain-specific SteelBERT model (Figure 13b) (MAE = 34.45 MPa, $R^2$ = 0.858) is comparable to Deepseek-R1 (Figure 13d) (MAE = 34.62 MPa, $R^2$ = 0.839), although SteelBERT for the particular data set shows slightly better agreement with experiment. A noteworthy trend is that more advanced or larger-scale models within the same family show better performance: for example, Deepseek-R1 compared to Deepseek-V3 (Figure 13c), similarly Gemini-2.5-Pro (Figure 13f) and Gemini-2.5-Flash (Figure 13e). Furthermore, the predictive accuracy of all LLMs is better than the baseline MLP model (Figure 13a) trained on traditional tabular data. Although these results are only for Al, they suggest that while domain-specific models could retain a marginal performance advantage, top-tier general-purpose LLMs appear to demonstrate comparable performance in capturing the intrinsic material properties through few-shot learning[134], without the need to separately use an MLP. Consequently, the performance of Deepseek-R1 and SteelBERT raises the following issue: As general-purpose LLMs become proficient at capturing the complex physical laws within specific scientific domains like materials science, the necessity of continuing to invest resources in developing and maintaining highly specialized, domain-specific models may warrant re-examination.

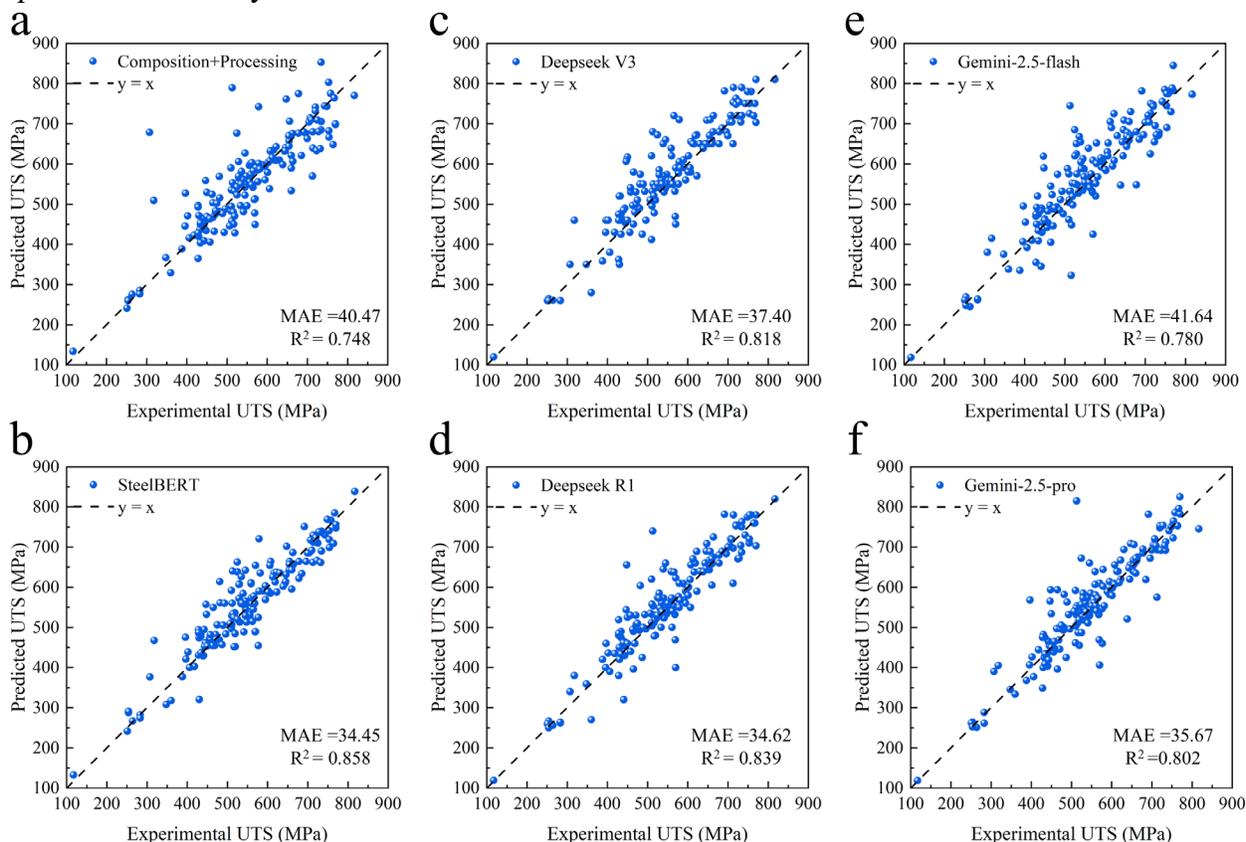

Figure 13. Performance of a baseline MLP and large language models on the aluminum alloy test set of Ref[135], a) An MLP model trained using tabular data of elemental compositions and processing conditions. It is compared with b) SteelBERT[100], a specially pretrained model for steels, c) Deepseek-V3, d) Deepseek-R1, e) Gemini-2.5-Flash, f) Gemini-2.5-Pro.

## 3.6 Considerations and Challenges in applying LLMs in Materials Science

We note that the superior predictive performance of SteelBERT in Ref[100] was achieved by extracting semantic embeddings of composition and processing and integrating them into a sophisticated hybrid downstream network. It is difficult to reproduce the results from such a hybrid network without a complete code. For example, if a Multi-Layer Perceptron (MLP) is used instead of the hybrid network, there is a significant disparity in performance. Figure 14a-c, employing a standard MLP as the downstream network for SciBERT, MatSciBERT, and SteelBERT, yields limited performance ($R^2 \approx 0.67 - 0.69$). While SteelBERT outperformed other BERT-based models because of its specialized pre-training on steel alloys, these findings suggest that although domain-specific pre-training is beneficial, fully leveraging its potential requires complex and suitable downstream networks which are equally critical for achieving optimal results. Hence, the question of whether it is possible to bypass domain-specific models and complex downstream networks altogether by directly leveraging the inherent reasoning capabilities of general-purpose LLMs. We tested this with Gemini on the training set of the same size, including compositions and processing parameters, as used in SteelBERT[100]. The results show that the predictive performance from Gemini-2.5-Flash (Figure 14d-f) is comparable to that of various BERT-based models such as SciBERT, MatSciBERT, and SteelBERT if utilizing a standard Multi-Layer Perceptron (MLP) as the downstream network. However, when employing the Gemini-2.5-Pro model, its prediction accuracy improves (Figure 14i) with results closely approaching those of the original SteelBERT study ($R^2=0.8256 \pm 0.0196$). Consistent with the previously discussed findings for aluminum alloys, this highlights that advanced general-purpose LLMs inherently possess robust few-shot learning capabilities.

Numerical experiments also suggest that data quantity imposes constraints on the effectiveness of fine-tuning. Although fine-tuning a model using methods, such as LoRA (Low Rank Adaptation), is generally considered an effective means of enhancing domain-specific performance, attempting to fine-tune even with a sizeable data set can yield no substantial improvement over the non-fine-tuned baseline. For example, this is seen with a data set of 983 points for high-entropy alloys using SteelBERT as a pretrained model. Now, SteelBERT is specifically trained on steel data and not on HEA data, even though HEA elements are well represented in the steel compositions. However, if instead of SteelBERT, we use Gemini or DeepSeek, the performance on the HEA data set is superior. Thus, recent LLMs, which incorporate aspects of sequential design using Reinforcement Learning, appear to use their "reasoning" capabilities to a greater degree in conjunction with data.

Learning potential relationships from existing data to predict the properties of new materials is an efficient approach; however, it requires both high quality data and insufficient quantity. Few-shot learning enables accurate predictions with limited samples by leveraging the knowledge embedded in pre-trained models. If we employ the retrieval-augmented generation (RAG) method to select the most relevant data from the training set, we find that when the selected samples account for less than 10% of the training data, model performance remains comparable to that achieved using the full dataset[134]. In one verification example, using only 2% of high-quality data, RAG+LLM can achieve $R^2 > 0.9$, indicating that data quality plays a significant role in guiding model inference. Under conditions of high data quality and limited data availability, a hybrid approach combining LLMs and RAG proves particularly effective. By screening optimal data and leveraging the reasoning capabilities of LLMs, highly accurate predictions can be obtained. This approach offers a feasible solution to the pervasive problem of data scarcity in materials science.

It is important to note the inference process of large language models (LLMs), slight variations in output may still occur even under identical few-shot prompting conditions with the temperature parameter set to zero (see Figure 14d-i). In an experiment involving 136 data points of the steel data set tested three times each, we found that Gemini-2.5-Flash produced identical outputs for 122 samples and non-identical outputs for 14 samples. Of the 14 samples, two of them had 2 identical outputs (Figure 14d-f). In contrast, Gemini-2.5-Pro yielded identical outputs for only 34 samples and non-identical for 102, with 60 of these showing two identical results. These findings indicate that the more "capable" or prone to less error the model is, the greater its output uncertainty. This has been observed in the context of reasoning that make LLM models more prone to hallucinations[136].

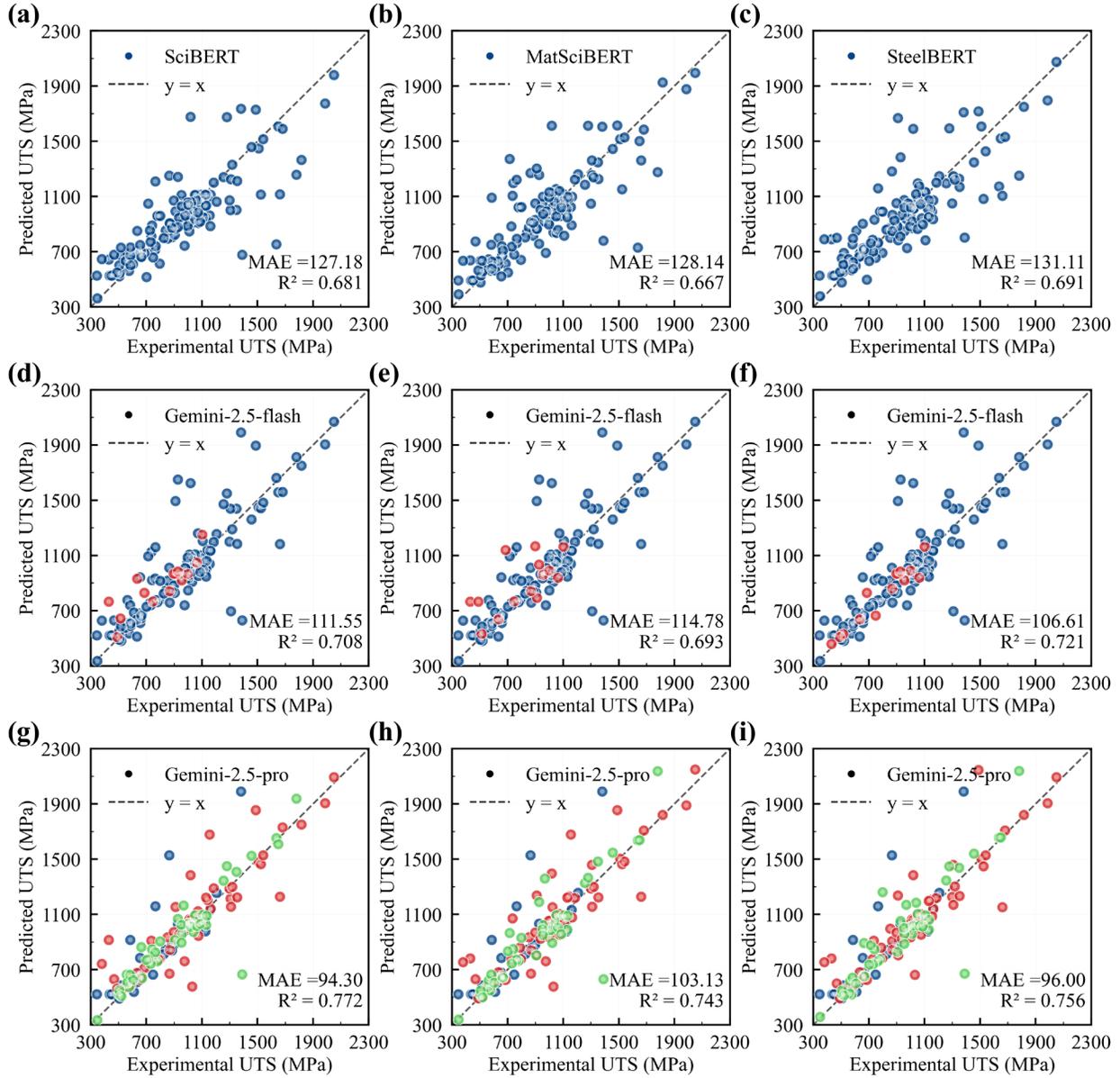

Figure 14 Performance comparison of various models in predicting the ultimate tensile strength (UTS) of the steel dataset[134]. a-c) shows the performance of a multilayer perceptron (MLP) model utilizing embeddings generated by SciBERT[137], MatSciBERT[110] and SteelBERT. d-f) illustrate three independent runs of Gemini-2.5-Flash, g-i) depict three independent runs of Gemini-2.5-Pro. For the Gemini models, data points are color-coded based on their run-to-run consistency: blue represents points identical across all three runs, red indicates points identical in exactly two runs, and green marks points that differ among all three runs.

The uncertainty primarily arises from non-deterministic computations during inference, including floating-point rounding errors, variations in distributed parallelism, fluctuations in cache states, and minor perturbations at tokenization boundaries. For example, the TPU-based mixed-precision architecture used by Gemini introduces higher numerical non-determinism in floating-point operations and retains small random perturbations by default to prevent output-mode collapse. Similarly, DeepSeek employs tensor parallelism and predictive decoding mechanisms, which inherently introduce slight randomness and quantization noise. Consequently, even with the temperature set to zero, these models cannot achieve fully deterministic outputs. Instead, they maintain probabilistic consistency, meaning that their outputs remain stable in distribution rather than identical token by token. Thus, in materials science the inherent randomness of LLM outputs is generally acceptable when the objective is qualitative understanding or theoretical insight. Minor variations in the model's responses do not substantially influence the explanatory power of trend analyses or conceptual reasoning. However, this uncertainty becomes problematic in quantitative prediction tasks, such as assessing thermodynamic stability or predicting mechanical properties. Even small inference perturbations may result in inconsistencies in predicted values or material recommendations, thereby compromising the reproducibility and reliability of results. To mitigate these issues, researchers should stabilize model performance by standardizing the inference environment, averaging results, or employing unified prompts to ensure that LLM-based research meets the rigor and verifiability required for scientific inquiry.

## 4 Outlook: Toward the Virtual Materials Scientist

Although deep neural networks demonstrate powerful predictive capabilities in materials science, their "black box" nature presents a core challenge. The mechanism by which their vast parameters cooperate to produce a final solution remains opaque, hindering the potential for AI to facilitate the discovery of new scientific knowledge. To address this, the academic community is actively exploring methods to enhance AI explainability. For example, SHAP (SHapley Additive exPlanations) analysis, rooted in cooperative game theory[138], quantifies each input feature's average marginal contribution to a model's prediction by considering all possible combinations or coalitions of features. While it accounts for pairwise (and higher-order) interactions—yielding intuitive force plots or dependence diagrams—SHAP is correlational, offering no insight into causal directionality or mechanisms (e.g., does composition *cause* strength, or vice versa?)[139]. Nevertheless, causal modeling and analysis do offer opportunities in materials science via the use of the PC algorithm (Peter-Clark)[140], a constraint-based method that infers directed acyclic graphs from observational data via conditional independence tests (e.g., Fisher's Z or mutual information). It can disentangle conflicting factors in alloy design, for example, by distinguishing the causal effect of elemental doping from heat-treatment processing on yield strength. Concurrently, more disruptive frameworks are emerging. Physics-Informed Neural Networks (PINNs)[141] embed physical laws directly into the training process, ensuring their outputs conform to established scientific principles. Meanwhile, Kolmogorov-Arnold Networks (KANs)[142,143], with their unique architecture, promise to simplify complex model relationships into intuitive mathematical formulas. These efforts to enhance model transparency and trustworthiness are laying the foundations of trust for a new, more intelligent paradigm in scientific research.

Machine Learning and Large Language Models (LLMs) are introducing unprecedented opportunities in materials science, with their influence expected to span the entire research workflow, from theoretical prediction to experimental validation. With advances in active learning, reinforcement learning, few-shot learning, uncertainty quantification, data security, inverse design, and integrated frameworks, we anticipate the emergence of an AI-driven paradigm for efficient, intelligent, and automated materials science. We are now witnessing LLMs undergoing considerable transformation, evolving from a tool with initial potential for machine translation and language summarization to a core component that is reshaping how materials science is being studied. We have already seen in Section 2.5 some evidence to suggest that pretraining an LLM with specialized data for certain domains does not necessarily yield better performance compared to a general LLM that has better reasoning capabilities. This is already a departure in our thinking from only a year or so ago. This trajectory points beyond optimizing performance on purely single tasks but rather toward the development of a "virtual scientist"—an intelligent, trustworthy, and autonomous agent deeply integrated into the entire research process, capable of independently designing experiments without necessarily depending on pretrained LLMs on specialized knowledge.

A fundamental challenge for LLMs, which operate by predicting the next word based on preceding context, is their inherent uncertainty. To transition from simple "prediction" to "reliable prediction," methods such as Retrieval-Augmented Generation (RAG) offer an effective solution. A dual strategy is essential for data utilization, addressing both the sparsity of "small data" and the complexity of "big data." For the former, prompt engineering techniques such as few-shot prompt and Chain of Thought (CoT) are playing a key role in unlocking the model's reasoning under data constraints. For large historical datasets, an "intelligent data" strategy that combines expert insights with RAG becomes central. Directly inputting massive datasets into a model is not only cost effective but also risks "catastrophic forgetting." A more forward-thinking proposal would be to create a domain-specific knowledge base where experts provide inferential annotations for data samples—analyzing and developing the causal links between composition, processing, and properties. Using RAG, the model can then retrieve a few highly relevant samples and their associated expert reasoning chains for new tasks. This approach offers a dual benefit: it guides the model to emulate expert thinking, yielding results that are both accurate and physically interpretable, while also transforming a complex "big data" problem into an efficient "few-shot prompt" task. This effectively can mitigate high computational costs and model memory limitations, providing a key step in evolving AI from a mere computational tool into a genuinely trustworthy research partner.

Given the nature of materials R&D data, a multi-agent system architecture is emerging with emphasis on collaboration, combining localized small models, which operate within a secure internal network to protect private data, with cloud-based large language models that provide extensive knowledge and superior reasoning capabilities. This architecture strikes a critical balance between security and access to state-of-the-art performance. This necessitates a decisive shift in the materials research mindset, moving from traditional "forward design"—a trial-and-error approach—to an "inverse design" paradigm driven by target properties. By integrating LLMs with methods such as Reinforcement Learning (RL), AI agents can efficiently conduct targeted explorations within the vast chemical space, significantly accelerating the process of designing new materials on demand.

Ultimately, these developments will foster integrated scientific discovery platforms that seamlessly connect the entire research chain, from data analysis and hypothesis generation to virtual screening, experimental design, and the automation of physical experiments. Within this framework, a new human-AI collaborative relationship will form. The AI system will assume the burden of processing massive amounts of information and exploring complex solution spaces, thereby liberating human scientists to concentrate on higher-level creative activities, such as strategic planning and identifying key research directions.

## 5 Acknowledgements

We acknowledge discussions on aspects of this work with colleagues at University of Beijing Science and Technology, Shanghai Jiatong University and Shanghai University. We are grateful for support from the National Natural Science Foundation of China (No. 52350710205, No. 52250191) and the Fundamental Research Funds for the Central Universities. This work is sponsored by the Key Research and Development Program of the Ministry of Science and Technology (No.2023YFB4604100). Support of the HPC Platform, Xi'an Jiaotong University is also acknowledged.